\documentclass[lettersize,journal]{IEEEtran}
\usepackage{amsmath,amsfonts}
\usepackage{algorithmic}
\usepackage{algorithm}
\usepackage{array}
\usepackage[caption=false,font=normalsize,labelfont=sf,textfont=sf]{subfig}
\usepackage{textcomp}
\usepackage{stfloats}
\usepackage{url}
\usepackage{verbatim}
\usepackage{graphicx}
\usepackage{cite}
\usepackage{booktabs}
\usepackage{amsmath}

\usepackage{xcolor}
\usepackage{amsmath,amssymb,amsthm}
\newtheorem{theorem}{Theorem}
\newtheorem{lemma}{Lemma}
\newtheorem{definition}{Definition}

\begin{document}

\bstctlcite{IEEEcontrol}

\title{Impact of Reduced-Order Modelling on Sum-of-Squares Region-of-Attraction Certification for Grid-Forming and Grid-Following Converters}

\author{Chenxi Yuan, Jeremy D. Watson, Neville R. Watson

\thanks{The authors are with the Department of Electrical and Computer Engineering, University of Canterbury, Christchurch, New Zealand (e-mail: chenxi.yuan@pg.canterbury.ac.nz; jeremy.watson@canterbury.ac.nz; neville.watson@canterbury.ac.nz).}%
}
\maketitle

\begin{abstract}
With the increasing penetration of power-electronic equipment, the transient stability of grid-forming (GFM) and grid-following (GFL) converters is increasingly characterized  through the region of attraction (ROA) of the post-fault equilibrium. Sum-of-squares (SOS) programming yields certified inner approximations of the ROA, yet its computational cost grows rapidly with system order. Reduced-order models are therefore widely used to reduce computational cost and, more importantly, to enable analytical investigation. However, ROA analyses based on these models often assume that reproducing the converter's transient response, or retaining more dynamics in the correct time-frame, is sufficient to preserve the ROA of the full-order system.
Using time-domain ODE simulations as a reference, the SOS-based ROA of each reduced-order GFM and GFL model is assessed against both its own simulated ROA and that of the corresponding full-order model, thereby separating SOS-induced conservativeness from model-order-induced ROA distortion.
The results indicate that increasing reduced-order model complexity does not necessarily improve SOS-based ROA fidelity. In the cases examined, more complex reduced-order models yield lower SOS coverage of the full-order reference ROA, despite closer ODE-boundary agreement in some regions. Model selection should therefore consider both full-order ROA fidelity and certified coverage of the practical operating range, rather than model complexity alone.
\end{abstract}


\section{Introduction}
\IEEEPARstart{T}{he} increasing penetration of power-electronic equipment is reshaping the dynamics of modern power systems. As synchronous generation is displaced and system strength declines, secure operation depends increasingly on the control dynamics of GFM and GFL converters \cite{9309174,10064337,10293188}. While classical small-signal analysis primarily characterizes system behavior under small deviations from an equilibrium point, converter-based power systems can experience severe disturbances, such as grid faults \cite{10106011} and switching operations \cite{9594094}. Assessing trajectory recovery under such disturbances is therefore a transient (large-signal) stability problem, for which linearization-based analysis is generally insufficient \cite{7404044}.

The ROA associated with a stable equilibrium point refers to the set of post-disturbance states from which the system trajectories asymptotically converge to that equilibrium point \cite{9497331}. Lyapunov-based ROA estimation supports the design and control of converter-based systems \cite{8428437,10310265,8036269}. Among existing ROA estimation methods, SOS programming is especially attractive, as it constructs the Lyapunov function automatically through constrained optimization and thereby attains a larger search space and lower conservativeness than fuzzy-model- or LMI-based approaches\cite{10310273}. The computational cost of SOS programming, however, grows rapidly with the number of states and the polynomial degree \cite{6469186}, making the certification of detailed full-order converter models impractical.

To ensure computational tractability, reduced-order converter models are widely adopted in ROA studies \cite{8444083,11141565}, obtained either through formal model-order-reduction techniques or through physically motivated time-scale separation that neglects fast inner-loop, filter, or PLL dynamics \cite{1425586,6901234,7974768,7933007}. Although reduced-order converter models are often validated against full-order models using transient waveforms \cite{6403906,9329082,10412691} or frequency-domain responses \cite{4349112,9570820}, such validation does not ensure that they provide an accurate representation of the full-order ROA. Conventional response-based comparisons assess model fidelity along selected trajectories or frequency ranges, whereas ROA analysis concerns the set of post-disturbance initial states that converge to the equilibrium point. Consequently, models with similar transient responses may still yield substantially different stability boundaries when their ROAs are compared on a common state-space slice, such as the \(\delta\)-\(\omega\) plane. This motivates a direct assessment of full and reduced-order ROAs for ROA-based stability analysis.

This paper investigates how reduced-order model selection affects SOS-based ROA certification for GFM and GFL converters. The main contributions are summarized as follows:
\begin{itemize}
    \item An ROA-oriented framework is developed to distinguish model-reduction error from SOS-certification conservatism. For each reduced-order model, its ODE-based ROA is first compared with that of the corresponding full-order model to assess ROA distortion due to model reduction, and its SOS-certified ROA is then compared with its own ODE-based ROA to quantify the additional conservatism introduced by SOS.

    \item The framework is applied systematically to typical GFM and GFL converter model hierarchies, comparing their transient-stability regions under consistent physical and operating conditions within each hierarchy. The results show that increased model order or improved dynamic fidelity does not necessarily yield a more accurate or more useful SOS-based ROA certificate.

    \item Practical guidance is provided for selecting reduced-order models for SOS-based ROA analysis, based on the combined requirements of full-order ROA fidelity, certified-region coverage and computational tractability, rather than model order alone.
\end{itemize}

The remainder of this paper is organized as follows. Section~II presents the ROA-based evaluation framework. Section~III introduces the GFM and GFL model families. Section~IV presents and compares ODE- and SOS-based ROA results. Section~V provides practical guidelines for model selection, and Section~VI concludes the paper.

\section{ROA-Oriented Benchmarking Formulation}
\subsection{SOS-Based Estimation of the ROA}
\label{subsec:sos_roa}

\subsubsection{Problem Formulation}

Consider an autonomous nonlinear dynamical system of the form
\begin{equation}
\dot{x}=f(x),
\qquad x\in\mathbb{R}^{n},
\qquad x(0)=x_{0},
\label{eq:sys}
\end{equation}
where $x$ is the state vector and $f:\mathbb{R}^{n}\to\mathbb{R}^{n}$ is a locally
Lipschitz continuous vector field. Without loss of generality, the equilibrium of
interest is assumed to be shifted to the origin, i.e., $f(0)=0$, which for the converter
models considered in this work is achieved by expressing the dynamics in terms of
deviations from the operating point.

\begin{definition}[Region of attraction]
Let the origin be an asymptotically stable equilibrium of \eqref{eq:sys}.
The region of attraction (ROA) is the set
\begin{equation}
\mathcal{R}
=\left\{x_{0}\in\mathbb{R}^{n}:
\lim_{t\to\infty}\phi(t,x_{0})=0\right\},
\label{eq:roa_def}
\end{equation}
where $\phi(t,x_{0})$ denotes the solution of \eqref{eq:sys} with initial condition
$x_{0}$.
\end{definition}

Since the exact ROA of a general nonlinear system does not admit a
closed-form characterization, the practical objective is to compute a
certified inner approximation of $\mathcal{R}$. The classical tool for this purpose is
Lyapunov's direct method.

\begin{theorem}[Lyapunov's direct method]
\label{thm:lyap}
Let $\mathcal{D}\subseteq\mathbb{R}^{n}$ be a domain containing the origin. If there
exists a continuously differentiable function
$V:\mathcal{D}\to\mathbb{R}$ such that
\begin{subequations}\label{eq:lyap_conditions}
\begin{align}
V(0)&=0,
\qquad
V(x)>0,
\quad
\forall x\in\mathcal{D}\setminus\{0\},
\label{eq:lyap_pos}\\
\dot{V}(x)
&=\nabla V(x)\cdot f(x)<0,
\quad
\forall x\in\mathcal{D}\setminus\{0\}.
\label{eq:lyap_dec}
\end{align}
\end{subequations}
then the equilibrium point is asymptotically stable.
\end{theorem}

\begin{lemma}[Invariant sublevel set]
\label{lem:sublevel}
Let $V$ satisfy the conditions of Theorem~\ref{thm:lyap} on
$\mathcal{D}$, and let $\lambda>0$ be such that the sublevel set
\begin{equation}
\Omega_{\lambda}
=\left\{x\in\mathbb{R}^{n}:V(x)\leq\lambda\right\}
\label{eq:sublevel}
\end{equation}
is bounded and contained in $\mathcal{D}$. Then $\Omega_{\lambda}$ is
positively invariant and $\Omega_{\lambda}\subseteq\mathcal{R}$, i.e., it provides a
certified inner approximation of the ROA \cite{4471858,5337881}.
\end{lemma}

The central challenge in applying Theorem~\ref{thm:lyap} is the
construction of a suitable Lyapunov function, which traditionally relies on experience or
on energy functions that may not exist for lossy systems. The SOS
technique removes this obstacle by transforming the search for $V(x)$ into a convex
semidefinite program (SDP).

\subsubsection{SOS Relaxation of the Lyapunov function}

\begin{definition}[SOS polynomial]
A polynomial $p(x)$ is a sum of squares (SOS), denoted $p\in\Sigma[x]$,
if there exist polynomials $q_{1},\dots,q_{m}$ such that
\begin{equation}
p(x)=\sum_{i=1}^{m}q_{i}^{2}(x).
\label{eq:sos_def}
\end{equation}
\end{definition}

Further details on SOS polynomials and their applications to polynomial
optimization and control can be found in\cite{parrilo2003semidefinite,5443730}.

Membership in $\Sigma[x]$ is a sufficient condition for global
non-negativity and is computationally tractable: $p\in\Sigma[x]$ if and
only if there exists $P\succeq0$ such that
\begin{equation}
p(x)=Z^{\top}(x)PZ(x),
\label{eq:gram}
\end{equation}
where $Z(x)$ is a vector of monomials up to half the degree of $p$. Finding such a $P$ is a semidefinite program, solvable via SOSTOOLS\cite{6070953,4505350} or YALMIP\cite{8439024,9865219} with standard SDP solvers. In this study, the resulting SDP is solved using MOSEK. Replacing the non-negativity conditions \eqref{eq:lyap_pos}--\eqref{eq:lyap_dec} by SOS constraints thus turns the construction of polynomial Lyapunov functions into a systematic optimization, at the cost of some conservatism since SOS is sufficient but not necessary for non-negativity. 

\subsubsection{ROA Estimation as an SOS Program}

Following the generalized $\mathcal{S}$-procedure, set-containment conditions between semialgebraic sets can be certified through SOS multipliers as sufficient conditions. Let $\rho(x)$ be a fixed, positive definite shape polynomial, typically $\rho(x)=x^{\top}x$, whose sublevel set $\mathcal{P}_{\beta}=\{x:\rho(x)\leq\beta\}$ provides a measure of the size of the certified region. The ROA estimation problem is formulated as the following SOS program \cite{CHAKRABORTY2011335,8959232}, which searches for a Lyapunov function while enlarging an inner approximation of the ROA:
\begin{equation}
\begin{aligned}
\max_{\substack{V,s_1,s_2\\ \lambda>0,\;\beta>0}}
\quad & \beta\\
\mathrm{s.t.}\quad
& V(0)=0,\\
& V(x)-L_1(x)\in\Sigma[x],\\
& \bigl(V(x)-\lambda\bigr)s_2(x)
-\nabla V(x)\cdot f(x)-L_2(x)
\in\Sigma[x],\\
& \bigl(\rho(x)-\beta\bigr)s_1(x)
-\bigl(V(x)-\lambda\bigr)
\in\Sigma[x],\\
& s_1,s_2\in\Sigma[x],\\
& L_1(x)=\varepsilon_1x^{\top}x,
\qquad L_2(x)=\varepsilon_2x^{\top}x,\\
& \varepsilon_1>0,\qquad\varepsilon_2>0.
\end{aligned}
\label{eq:sos_program}
\end{equation}
Here, $\varepsilon_1$ and $\varepsilon_2$ are prescribed positive constants. The positivity constraint enforces strict positive definiteness of $V$. The Lyapunov-decrease constraint ensures $\dot{V}<0$ on $\Omega_{\lambda}\setminus\{0\}$, where $\Omega_{\lambda}=\{x:V(x)\leq\lambda\}$. Together with Lemma~\ref{lem:sublevel}, this establishes $\Omega_{\lambda}\subseteq\mathcal{R}$. The shape-set containment constraint ensures $\mathcal{P}_{\beta}\subseteq\Omega_{\lambda}$. Thus, maximizing $\beta$ enlarges the shape set contained within the certified Lyapunov sublevel set.

For the converter models considered here, auxiliary state variables are introduced to replace the trigonometric terms, while their algebraic relation is retained as an equality constraint \cite{7163521,11426902,Zhang2020SOS}. The notation $x$ denotes the resulting state vector after equilibrium shifting and normalization. The auxiliary variables satisfy the algebraic constraint 
\begin{equation}
g(x)=S(x)^2+C(x)^2-1=0,
\label{eq:lifted_algebraic_constraint}
\end{equation}
where $S(x)$ and $C(x)$ express $\sin\delta$ and $\cos\delta$ in terms of the normalized coordinates $x$, and $\delta$ is the physical angle state.

The inequality $h(x)\geq0$ defines the admissible domain, enforcing a voltage lower bound for the GFM models and a positive lower bound on the PLL denominator for the GFL models. For the rational GFL models, multiplying the vector field by this positive denominator yields a polynomial vector field while preserving trajectory paths and their direction.

On the set defined by $g(x)=0$ and $h(x)\geq0$, the Lyapunov-decrease and shape-set containment conditions in \eqref{eq:sos_program} are expressed using polynomial multipliers \cite{Papachristodoulou2005,Zhang2020SOS} as
\begin{subequations}
\label{eq:lifted_sos_constraints}
\begin{align}
& (V-\lambda)s_2-\nabla V\cdot f-L_2
  \notag\\
& \qquad {}-\ell_2g-\sigma_2h\in\Sigma[x],
\label{eq:lifted_sos_decrease}\\
& (\rho-\beta)s_1-(V-\lambda)
  \notag\\
& \qquad {}-\ell_1g-\sigma_1h\in\Sigma[x].
\label{eq:lifted_sos_containment}
\end{align}
\end{subequations}
Here, $\ell_1$ and $\ell_2$ are free polynomial multipliers,
and $\sigma_1,\sigma_2\in\Sigma[x]$.

The Lyapunov sublevel set is required to lie within the admissible set \cite{6469186}. This inclusion is enforced by
\begin{equation}
h-s_3(\lambda-V)-\ell_3g\in\Sigma[x],
\label{eq:sos_domain_containment}
\end{equation}
where $s_3\in\Sigma[x]$ and $\ell_3$ is a free polynomial multiplier.

Define the algebraic set $\mathcal{M}=\{x:g(x)=0\}$ and the admissible set $\mathcal{D}=\{x\in\mathcal{M}:h(x)\geq0\}$. The Lyapunov and shape sublevel sets are $\Omega_{\lambda}=\{x\in\mathcal{M}:V(x)\leq\lambda\}$ and $\mathcal{P}_{\beta}=\{x\in\mathcal{D}:\rho(x)\leq\beta\}$, respectively. The SOS constraints certify $\mathcal{P}_{\beta}\subseteq\Omega_{\lambda} \subseteq\mathcal{D}$ and strict Lyapunov decrease on $\Omega_{\lambda}\setminus\{0\}$. The resulting ROA estimate in the transformed coordinates is 
\begin{equation}
\widehat{\mathcal{R}}
=\left\{x\in\mathcal{M}:
V^{\star}(x)<\lambda^{\star}\right\},
\label{eq:roa_estimate}
\end{equation}
where $V^{\star}$ and $\lambda^{\star}$ denote the final accepted solution. The estimate in the original physical coordinates is obtained by substituting the coordinate transformation into \eqref{eq:roa_estimate}.

\subsubsection{Solution Algorithm}

The SOS formulation is bilinear in the decision variables because of products such as $Vs_2$, $\beta s_1$, and $Vs_3$. An alternating scheme is therefore adopted \cite{CHAKRABORTY2011335,Zhang2020SOS}. The Lyapunov update is a convex SDP with the principal multipliers fixed, whereas the multiplier update uses scalar bisection with an SDP feasibility problem at each trial value of $\beta$. The specific implementation is detailed in Appendix~\ref{SOS_Algorithm}.

\subsection{Simulation-Based Reference ROA}

To evaluate both the conservatism of the SOS estimate $\widehat{\mathcal{R}}$ and the fidelity of the reduced-order models, a reference ROA, denoted $\mathcal{R}_{\mathrm{ref}}$, is computed for each model by exhaustive time-domain simulation: the nonlinear ODE model is integrated forward from a grid of initial conditions, and a grid point is classified as stable if its trajectory converges to the equilibrium within a prescribed tolerance over the simulation horizon.

Each trajectory is simulated for \(10\,\mathrm{s}\), with convergence assessed at all sampled points in the final \(2\,\mathrm{s}\). The angle and frequency deviations must remain within \(2^\circ\) and \(1\,\mathrm{rad/s}\), respectively. Additional voltage and current tolerances are \(5\,\mathrm{V}\) and \(0.5\,\mathrm{A}\) for the GFM models. For GFL-5/16, the current and squared per-unit DC-voltage tolerances are \(0.01\,\mathrm{pu}\) and \(0.02\), respectively. Model-specific tolerances also apply to the remaining internal states, and trajectories triggering termination events are excluded. These parameters and tolerances were chosen via testing a range of different values and inspecting results in the time-domain, confirming that the resulting ROA boundaries are accurately portrayed.

Since exhaustive gridding is tractable only in low dimensions, all comparisons are carried out on a common two-dimensional slice of the state space, with the remaining states initialized at their equilibrium values, so that the reduced- and full-order ROAs are compared on identical cross sections.

The reference ROA converges to the exact ROA as the grid is refined and the horizon extended, but is itself subject to discretization and finite simulation time; within these limits, it serves as the common reference for all comparisons.

\section{Representative Models \\ of GFM and GFL Converters}

\subsection{GFM Converter Models}

The parameters used for the three GFM benchmark models are summarized in Table~\ref{tab:gfm_parameters}. Identical operating, droop, network, and load parameters are employed to ensure a consistent comparison. The no-load frequency setpoint is adjusted according to \(\omega_n=\omega_0+k_pP^\ast\) to maintain \(\omega^\ast=\omega_0\) at the selected nonzero-power operating point, where \(P^\ast\) denotes the steady-state power entering the active-power droop equation.

\begin{table*}[!t]
\caption{System and Control Parameters of the GFM Benchmark Models}
\label{tab:gfm_parameters}
\centering
\renewcommand{\arraystretch}{1.15}
\setlength{\tabcolsep}{6pt}
\begin{tabular}{lccccc}
\hline
Parameter & Symbol & GFM-15 & GFM-5 & GFM-6 & Unit \\
\hline

\multicolumn{6}{c}{\textit{Common Operating and Droop Parameters}} \\
\hline
Nominal angular frequency
& $\omega_{0}$ & $100\pi$ & $100\pi$ & $100\pi$
& $\mathrm{rad/s}$ \\

Nominal voltage set point
& $V_{n}$ & $311$ & $311$ & $311$
& $\mathrm{V}$ \\

Stiff-grid voltage
& $V_{0}$ & $311$ & $311$ & $311$
& $\mathrm{V}$ \\

Active-power droop coefficient
& $k_{p}$ & $9.4\times10^{-5}$ & $9.4\times10^{-5}$
& $9.4\times10^{-5}$
& $\mathrm{rad/(s\,W)}$ \\

Reactive-power droop coefficient
& $k_{q}$ & $1.0\times10^{-4}$ & $1.0\times10^{-4}$
& $1.0\times10^{-4}$
& $\mathrm{V/VAr}$ \\

Power-filter time constant
& $\tau$ & $3.18\times10^{-2}$ & $3.18\times10^{-2}$
& $3.18\times10^{-2}$
& $\mathrm{s}$ \\

\hline
\multicolumn{6}{c}{\textit{Network and Load Parameters}} \\
\hline
Grid-line resistance
& $R_{\ell}$ & $0.195$ & $0.195$ & $0.195$
& $\Omega$ \\

Grid-line inductance
& $L_{\ell}$ & $6.1\times10^{-4}$ & $6.1\times10^{-4}$
& $6.1\times10^{-4}$
& $\mathrm{H}$ \\

Load resistance
& $R_{L}$ & $20$ & $20$ & $20$
& $\Omega$ \\

Load inductance
& $L_{L}$ & $1.5\times10^{-2}$ & $1.5\times10^{-2}$
& $1.5\times10^{-2}$
& $\mathrm{H}$ \\

\hline
\multicolumn{6}{c}{\textit{Filter and Inner-Control Parameters}} \\
\hline
Filter resistance
& $R_{f}$ & $0.05$ & -- & --
& $\Omega$ \\

Filter inductance
& $L_{f}$ & $8.0\times10^{-3}$ & -- & --
& $\mathrm{H}$ \\

Filter capacitance
& $C_{f}$ & $150\times10^{-6}$ & -- & $150\times10^{-6}$
& $\mathrm{F}$ \\

Voltage-loop proportional gain
& $K_{PV}$ & $5$ & -- & --
& $\mathrm{A/V}$ \\

Voltage-loop integral gain
& $K_{IV}$ & $100$ & -- & --
& $\mathrm{A/(V\,s)}$ \\

Current-loop proportional gain
& $K_{PC}$ & $5$ & -- & --
& $\mathrm{V/A}$ \\
 
Current-loop integral gain
& $K_{IC}$ & $100$ & -- & --
& $\mathrm{V/(A\,s)}$ \\

Equivalent voltage-control gain
& $G_{v}$ & -- & -- & $5$
& $\mathrm{A/V}$ \\
\hline
\end{tabular}
\end{table*}

\subsubsection{Full-Order Model (GFM-15)}
\label{subsubsec:gfm15}

The full-order grid-forming benchmark is a continuous-time average
model of a droop-controlled voltage-source inverter connected to a
stiff grid \cite{4118327,ojo2020reviewreducedordermodelsmicrogrids}.
It retains the frequency and voltage droop dynamics, outer voltage
controller, inner current controller, $LC$-filter dynamics, grid-line
dynamics, and local-load dynamics. The dc source is assumed to have
sufficient energy reserves, and the switching process is represented
by its average model. 

The state vector is
\(\boldsymbol{x}_{15}=[
\delta,\allowbreak \omega,\allowbreak V,\allowbreak
\boldsymbol{\phi}_{dq}^{\mathsf T},\allowbreak
\boldsymbol{\gamma}_{dq}^{\mathsf T},\allowbreak
\boldsymbol{i}_{dq}^{\mathsf T},\allowbreak
\boldsymbol{v}_{odq}^{\mathsf T},\allowbreak
(\boldsymbol{I}_{DQ}^{\ell})^{\mathsf T},\allowbreak
(\boldsymbol{I}_{DQ}^{L})^{\mathsf T}
]^{\mathsf T}\in\mathbb{R}^{15}\).
Its nonlinear dynamics are written compactly as \(\dot{\boldsymbol{x}}_{15}
=\boldsymbol{f}_{15}(\boldsymbol{x}_{15})\).

The complete state equations are provided in
Appendix~\ref{app:gfm_full_model}. In the ROA calculations, the
equilibrium of interest is shifted to the origin using
$\widetilde{\boldsymbol{x}}_{15}
=\boldsymbol{x}_{15}-\boldsymbol{x}_{15}^{\star}$.
The GFM-15 model is used only as the full-order simulation benchmark;
the subsequent reduced-order models provide the polynomial dynamics
used in the SOS analysis.

\subsubsection{Reduced-Order Model (GFM-5)}

Following the conventional third-order GFM representation in
\cite{ojo2020reviewreducedordermodelsmicrogrids}, the inner control loops and $LC$-filter dynamics are neglected by assuming ideal voltage tracking, such that
$v_{odq}=eV$. The line dynamics are further replaced by their
quasi-steady-state approximation, yielding
\begin{subequations}\label{eq:gfm5_line_power}
\begin{align}
P_{\ell}
&=GV^{2}-GVV_{0}\cos\delta+BVV_{0}\sin\delta,
\label{eq:gfm5_line_active_power}\\
Q_{\ell}
&=BV^{2}-BVV_{0}\cos\delta-GVV_{0}\sin\delta.
\label{eq:gfm5_line_reactive_power}
\end{align}
\end{subequations}
where
\begin{subequations}\label{eq:gfm5_admittance}
\begin{align}
G&=\frac{R_{\ell}}{R_{\ell}^{2}+X_{\ell}^{2}},
\label{eq:gfm5_conductance}\\
B&=\frac{X_{\ell}}{R_{\ell}^{2}+X_{\ell}^{2}},
\label{eq:gfm5_susceptance}\\
X_{\ell}&=\omega_{0}L_{\ell}.
\label{eq:gfm5_reactance}
\end{align}
\end{subequations}

Retaining the dynamic $RL$ load as in \cite{ojo2020reviewreducedordermodelsmicrogrids}, the reduced-order model is
\begin{subequations}\label{eq:gfm5_model}
\begin{align}
\dot{\delta}&=\omega-\omega_{0},
\label{eq:gfm5_delta}\\
\tau\dot{\omega}
&=-\omega+\omega_{n}-k_{p}(P_{L}+P_{\ell}),
\label{eq:gfm5_omega}\\
\tau\dot{V}
&=-V+V_{n}-k_{q}(Q_{L}+Q_{\ell}),
\label{eq:gfm5_voltage}\\
L_{L}\dot{I}_{DQ}^{L}
&=\left(-R_{L}I_{2}+\omega_{0}L_{L}J\right)I_{DQ}^{L}
+\mathcal{T}(\delta)eV,
\label{eq:gfm5_load_dynamics}
\end{align}
\end{subequations}
where
\begin{subequations}\label{eq:gfm5_load_power}
\begin{align}
P_{L}
&=(I_{DQ}^{L})^{\top}\mathcal{T}(\delta)eV,
\label{eq:gfm5_load_active_power}\\
Q_{L}
&=(I_{DQ}^{L})^{\top}J\mathcal{T}(\delta)eV.
\label{eq:gfm5_load_reactive_power}
\end{align}
\end{subequations}

With the dynamic load retained, the resulting model is denoted GFM-5 and has the state vector with three converter states and two load states \(x_{5}=\bigl[\delta,\omega,V,I_{D}^{L},I_{Q}^{L}\bigr]^{\top}\). 

GFM-5 and GFM-15 share the same equilibrium values of the retained states. Their ROAs can therefore be compared around a common operating point, allowing the effect of model-order reduction to be assessed.

\subsubsection{Enhanced Reduced-Order Model (GFM-6)}

To improve the representation of the voltage dynamics, GFM-5 is extended by introducing an additional state $x_v$, which represents the combined effects of the inner control loops and filter-inductor dynamics. This state is driven by the voltage-droop error, $\tau\dot{x}_v=V_n-V-k_qQ$, and is coupled to the terminal-voltage dynamics through the term $G_v(x_v-V)$. A frequency-response comparison with GFM-15 and GFM-5 is provided in Appendix~\ref{app:gfm6_frequency} to support this model extension.

The static line model and dynamic $RL$ load are identical to those of GFM-5. Defining
\begin{subequations}\label{eq:gfm6_total_power}
\begin{align}
P&=P_{L}+P_{\ell},
\label{eq:gfm6_total_active_power}\\
Q&=Q_{L}+Q_{\ell},
\label{eq:gfm6_total_reactive_power}
\end{align}
\end{subequations}
the resulting model is
\begin{subequations}\label{eq:gfm6_model}
\begin{align}
\dot{\delta}&=\omega-\omega_{0},
\label{eq:gfm6_delta}\\
\tau\dot{\omega}&=-\omega+\omega_{n}-k_{p}P,
\label{eq:gfm6_omega}\\
C_{f}\dot{V}&=-\frac{Q}{V_{0}}+G_{v}(x_{v}-V),
\label{eq:gfm6_voltage}\\
\tau\dot{x}_{v}&=-V+V_{n}-k_{q}Q,
\label{eq:gfm6_xv}\\
L_{L}\dot{I}_{DQ}^{L}
&=\left(-R_{L}I_{2}+\omega_{0}L_{L}J\right)I_{DQ}^{L}
+\mathcal{T}(\delta)eV,
\label{eq:gfm6_load_dynamics}
\end{align}
\end{subequations}
where $G_{v}$ is an aggregate voltage-control parameter and $V_{0}$
denotes the nominal voltage used in the reactive-power normalization.

The complete model has six states,
\begin{equation}
x_{6}=\bigl[\delta,\omega,V,x_{v},I_{D}^{L},I_{Q}^{L}\bigr]^{\top},
\label{eq:gfm6_states}
\end{equation}
and is therefore denoted GFM-6. Its equilibrium values for the states
shared with GFM-5 and GFM-15 remain unchanged, while the additional
state satisfies
\begin{equation}
x_{v}^{\star}=V^{\star}+\frac{Q^{\star}}{G_{v}V_{0}}.
\label{eq:gfm6_xv_equilibrium}
\end{equation}
As $G_{v}\rightarrow\infty$, the additional voltage dynamics approach
the ideal voltage-tracking assumption, and GFM-6 reduces to GFM-5.

\subsection{GFL Converter Models}
\label{subsec:gfl_models}

Three GFL models are considered: the 16-state full-order model
(GFL-16), a reparameterized literature-based five-state model
(GFL-5), and a conventional three-state model (GFL-3). The same
physical parameters and operating commands are used wherever the
individual model structures permit. Each model is solved for its own
equilibrium before its ROA section is expressed in centered
coordinates.

The parameters used for the three GFL benchmark models are summarized in Table~\ref{tab:gfl_parameters}. Because GFL-5 is expressed in per-unit variables and GFL-3 uses normalized PLL coefficients, the corresponding mapped parameters are reported separately in Table~\ref{tab:gfl_parameters}.

\begin{table*}[!t]
\caption{System and Control Parameters of the GFL Benchmark Models}
\label{tab:gfl_parameters}
\centering
\footnotesize
\renewcommand{\arraystretch}{1.10}
\setlength{\tabcolsep}{3.5pt}
\begin{tabular}{lccccc}
\hline
Parameter & Symbol & GFL-16 & GFL-5 & GFL-3 & Unit \\
\hline

\multicolumn{6}{c}{\textit{Common Operating Parameters}} \\
\hline
Nominal angular frequency
& $\omega_0$ & $100\pi$ & $100\pi$ & $100\pi$
& $\mathrm{rad/s}$ \\

Grid-voltage magnitude
& $V_g$ & $326.599$ & $326.599$ & $326.599$
& $\mathrm{V}$ \\

DC-link voltage reference
& $V_{\rm dc}^{\rm ref}$ & $680$ & $680^{\dagger}$ & --
& $\mathrm{V}$ \\

DC-source current
& $I_{\rm src}$ & $100$ & -- & --
& $\mathrm{A}$ \\

Active-power operating point
& $P^{\star}$ & $68$ & $68^{\dagger}$ & $68$
& $\mathrm{kW}$ \\

Reactive-power operating point
& $Q^{\star}$ & $0$ & $0$ & $0$
& $\mathrm{kVAr}$ \\

\hline
\multicolumn{6}{c}{\textit{Electrical-Circuit Parameters}} \\
\hline
Converter-side inductance
& $L_1$ & $6.1\times10^{-4}$ & -- & --
& $\mathrm{H}$ \\

Grid-side inductance
& $L_2$ or $L_g$ & $6.1\times10^{-4}$
& $(6.1\times10^{-4})^{\dagger}$ & $6.1\times10^{-4}$
& $\mathrm{H}$ \\

Filter capacitance
& $C_f$ & $1.5\times10^{-4}$ & -- & --
& $\mathrm{F}$ \\

DC-link capacitance
& $C_{\rm dc}$ & $7.8\times10^{-2}$
& $(7.8\times10^{-2})^{\dagger}$ & --
& $\mathrm{F}$ \\

\hline
\multicolumn{6}{c}{\textit{Measurement and PWM Parameters}} \\
\hline
PWM-delay time constant
& $T_d$ & $5.0\times10^{-6}$ & -- & --
& $\mathrm{s}$ \\

Reactive-power measurement time constant
& $T_m$ & $3.18\times10^{-2}$ & -- & --
& $\mathrm{s}$ \\

Active-power-state time constant
& $\tau_P$ & -- & -- & $3.18\times10^{-2}$
& $\mathrm{s}$ \\

\hline
\multicolumn{6}{c}{\textit{PLL and Control Parameters}} \\
\hline
PLL proportional coefficient
& $K_{p,{\rm pll}}$ & $50$ & $50^{\dagger}$ & $50^{\ddagger}$
& $\mathrm{s^{-1}}$ \\

PLL integral coefficient
& $K_{i,{\rm pll}}$ & $250$ & $250^{\dagger}$ & $250^{\ddagger}$
& $\mathrm{s^{-2}}$ \\

Current-loop proportional gain
& $K_{pi}$ & $5$ & -- & --
& $\mathrm{kV/kA}$ \\

Current-loop integral time constant
& $T_{ii}$ & $1\times10^{-2}$ & -- & --
& $\mathrm{s}$ \\

DC-voltage-loop proportional gain
& $K_{pv}$ & $0.72$ & $0.72^{\dagger}$ & --
& $\mathrm{kA/kV}$ \\

DC-voltage-loop integral time constant
& $T_{iv}$ & $1.7\times10^{-2}$
& $(1.7\times10^{-2})^{\dagger}$ & --
& $\mathrm{s}$ \\

Reactive-loop proportional gain
& $K_{pq}$ & $0.3$ & $0.3^{\dagger}$ & --
& $\mathrm{kA/MVAr}$ \\

Reactive-loop integral time constant
& $T_{iq}$ & $2.0\times10^{-2}$
& $(2.0\times10^{-2})^{\dagger}$ & --
& $\mathrm{s}$ \\
\hline
\end{tabular}

\vspace{1mm}
\parbox{0.98\textwidth}{\footnotesize
$^{\dagger}$The marked physical and controller quantities are not inserted
directly into the GFL-5 equations. They are converted to the per-unit
PLL--DVC--TVC parameters listed in
Table~\ref{tab:gfl_parameters}.
$^{\ddagger}$GFL-3 uses the normalized gains
$k_p=K_{p,{\rm pll}}/V_g$ and
$k_i=K_{i,{\rm pll}}/V_g$, rather than using
$K_{p,{\rm pll}}$ and $K_{i,{\rm pll}}$ directly.}
\end{table*}

\subsubsection{Full-Order Model (GFL-16)}
\label{subsubsec:gfl16}

The full-order grid-following benchmark retains the LCL filter, dc-link dynamics, dc-voltage and reactive-power controllers, inner current controller, PWM delay, reactive-power measurement filter, and PLL dynamics. It is adopted from \cite{D.Maheepala} and written using the positive relative-angle convention
\begin{equation}
\delta_{16}=\theta_g-\theta_{\rm pll},
\qquad
\Delta\omega_{16}=\dot{\delta}_{16}
=\omega_g-\omega_{\rm pll}.
\label{eq:gfl16_angle_definition}
\end{equation}
The state vector is
\(\boldsymbol{x}_{16}=[
i_{1D},\allowbreak i_{1Q},\allowbreak
i_{2D},\allowbreak i_{2Q},\allowbreak
v_{cD},\allowbreak v_{cQ},\allowbreak
V_{\rm dc},\allowbreak
\xi_{id},\allowbreak \xi_{iq},\allowbreak
\xi_v,\allowbreak \xi_q,\allowbreak
\delta_{16},\allowbreak \xi_{\rm pll},\allowbreak
E_D,\allowbreak E_Q,\allowbreak Q_f
]^{\mathsf T}\in\mathbb{R}^{16}\).
Its nonlinear dynamics are represented compactly by
\(\dot{\boldsymbol{x}}_{16}
=\boldsymbol{f}_{16}(\boldsymbol{x}_{16})\).

The complete state equations are given in
Appendix~\ref{app:gfl_full_model}. For the ROA analysis, GFL-16 is
centered at its independently calculated equilibrium, such that
$\widetilde{\boldsymbol{x}}_{16}
=\boldsymbol{x}_{16}-\boldsymbol{x}_{16}^{\star}$. Its
simulation-based ROA is used as the full-order reference when
assessing the reduced GFL models.

\subsubsection{Reparameterized Five-State Model (GFL-5)}
\label{subsubsec:gfl5}

GFL-5 adopts the five-state outer-loop model structure presented in \cite{zhang2026bandwidth}. The inner current loop is represented as an ideal controlled current source, whereas the PLL, dc-link voltage controller (DVC), and terminal-voltage controller (TVC) are retained. The GFL-5 model is reparametrized using the common benchmark data for GFL-16 and GFL-3.

The state vector is
\(\boldsymbol{x}_{5}
=[\delta_{5},x_{\rm int,pll},i_d,z,i_q]^{\mathsf T}\),
where \(z=v_{\rm dc}^{2}-v_{\rm dc,ref}^{2}\). All electrical quantities in this model are expressed in per unit, whereas time is retained in seconds. The model retains the native relative-angle coordinate $\delta_5$ used by its source formulation.

Define the PLL denominator, ac-side output power, and power imbalance
as
\begin{subequations}\label{eq:gfl5_auxiliary}
\begin{align}
D_{\rm pll}
&=1-k_{p,\rm pll}\frac{X_g}{\omega_s}i_d,
\label{eq:gfl5_denominator}\\
p
&=i_dU_g\cos\delta_{5}
-i_qU_g\sin\delta_{5},
\label{eq:gfl5_power}\\
e_P&=P_{\rm in}-p.
\label{eq:gfl5_power_error}
\end{align}
\end{subequations}
Here, $X_g$ denotes the per-unit grid reactance. Its dynamic-inductance
coefficient satisfies
\begin{equation}
\frac{X_g}{\omega_s}
=\frac{L_g^{\rm SI}}{Z_{\rm base}},
\label{eq:gfl5_inductance_conversion}
\end{equation}
which is the coefficient multiplying $i_d\dot{\delta}_5$ after
conversion from SI quantities to per unit.

The five differential equations implemented in the benchmark are
\begin{subequations}\label{eq:gfl5_dynamics}
\begin{align}
\dot\delta_{5}
&=
\frac{\omega_{\rm pll}}{D_{\rm pll}}
\left(
X_gi_d-U_g\sin\delta_{5}
+\frac{x_{\rm int,pll}}{\omega_{\rm pll}}
\right),
\label{eq:gfl5_delta}\\
\dot x_{\rm int,pll}
&=
\frac{\omega_{\rm pll}\zeta_{\rm pll}}{D_{\rm pll}}
\left(
X_gi_d-U_g\sin\delta_{5}
+x_{\rm int,pll}i_d\frac{X_g}{\omega_s}
\right),
\label{eq:gfl5_pll_integrator}\\
\dot i_d
&=\omega_{\rm dvc}
\left[
e_P+
\frac{C_{\rm dc}}{2\omega_s}
\zeta_{\rm dvc}\omega_{\rm dvc}z
\right],
\label{eq:gfl5_id}\\
\dot z
&=\frac{2\omega_s}{C_{\rm dc}}e_P,
\label{eq:gfl5_z}\\
\dot i_q
&=\omega_{\rm tvc}
\left[
-i_q+k_vU_g\cos\delta_{5}
-k_vX_gi_q-k_vV_{t,\rm ref}
\right].
\label{eq:gfl5_iq}
\end{align}
\end{subequations}
The dc-link voltage represented by $z$ is recovered as
\begin{equation}
V_{\rm dc}
=V_{\rm dc,base}
\sqrt{v_{\rm dc,ref}^{2}+z}.
\label{eq:gfl5_vdc_recovery}
\end{equation}

For parameter consistency, the bases used by GFL-5 are selected as
\begin{subequations}\label{eq:gfl5_parameter_bases}
\begin{align}
V_{\rm base}&=V_g,
&S_{\rm base}&=P_{\rm ref},
&I_{\rm base}&=\frac{S_{\rm base}}{1.5V_{\rm base}},
\label{eq:gfl5_ac_bases}\\
Z_{\rm base}&=\frac{V_{\rm base}}{I_{\rm base}},
&V_{\rm dc,base}&=V_{\rm dc}^{\rm ref}.
\label{eq:gfl5_impedance_base}
\end{align}
\end{subequations}
Consequently, $U_g=P_{\rm in}=v_{\rm dc,ref}=1$ in the
per-unit model, while
\begin{equation}
X_g=\frac{\omega_sL_2}{Z_{\rm base}},
\qquad
C_{\rm dc}
=\frac{\omega_s C_{\rm dc}^{\rm SI}V_{\rm dc,base}^{2}}
{S_{\rm base}}.
\label{eq:gfl5_reparameterized_network}
\end{equation}
The common PLL gains are mapped according to
\begin{equation}
\omega_{\rm pll}=K_{p,\rm pll},
\qquad
\zeta_{\rm pll}
=\frac{K_{i,\rm pll}}{K_{p,\rm pll}}.
\label{eq:gfl5_pll_mapping}
\end{equation}
The common dc-voltage PI gains are converted to the corresponding
$\omega_{\rm dvc}$ and $\zeta_{\rm dvc}$ representation, whereas the
TVC parameters are selected from the common reactive-control data,
with $\omega_{\rm tvc}=1/T_{iq}$. Finally, $V_{t,\rm ref}$ is
recalculated so that the independently solved GFL-5 equilibrium
satisfies the common active-power command and zero grid reactive
power. These substitutions retain the five-state equation structure
of \cite{zhang2026bandwidth}, while placing GFL-5 at the operating and
parameter conditions used in the present comparison.

\subsubsection{Conventional Three-State Model (GFL-3)}
\label{subsubsec:gfl3}

GFL-3 is derived from the four-state model in \cite{10667279}, retaining the nonlinear PLL and a first-order active-power state. Its physical state vector is \(\boldsymbol{x}_3=
\begin{bmatrix}
\delta_3 & x_{\rm pll} & P
\end{bmatrix}^{\mathsf T}\).

In this formulation, $P$ is the peak-$dq$ power coefficient used by
the source model; consequently, the common three-phase power command
is mapped as $P^\star=(2/3)P_{\rm ref}$ in the implementation. For
compactness, define
\begin{subequations}\label{eq:gfl3_auxiliary}
\begin{align}
D_3(P)
&=1-\frac{k_pL_g}{V_g}P,
\label{eq:gfl3_denominator}\\
N_3(\delta_3,x_{\rm pll})
&=-k_pV_g\sin\delta_3+x_{\rm pll}.
\label{eq:gfl3_numerator}
\end{align}
\end{subequations}
The model equations are
\begin{subequations}\label{eq:gfl3_dynamics}
\begin{align}
\dot\delta_3
&=\frac{N_3}{D_3}-\omega_0,
\label{eq:gfl3_delta}\\
\dot x_{\rm pll}
&=\frac{k_i}{D_3}
\left[
-D_3V_g\sin\delta_3
+N_3\frac{L_gP}{V_g}
\right],
\label{eq:gfl3_pll}\\
\tau\dot P&=P^\star-P.
\label{eq:gfl3_power_state}
\end{align}
\end{subequations}
The stable operating point used to initialize the equilibrium solve is
\begin{subequations}\label{eq:gfl3_equilibrium}
\begin{align}
P_3^\star&=P^\star,
\label{eq:gfl3_Peq}\\
\delta_3^\star
&=\sin^{-1}\!\left(
\frac{\omega_0L_gP^\star}{V_g^2}
\right),
\label{eq:gfl3_deltaeq}\\
x_{\rm pll}^\star&=\omega_0.
\label{eq:gfl3_xplleq}
\end{align}
\end{subequations}
In the numerical implementation, the integrated states are the
deviations $\delta_3-\delta_3^\star$,
$x_{\rm pll}-x_{\rm pll}^\star$, and $P-P_3^\star$. Compared with
GFL-16, GFL-3 omits the dc-link voltage, DVC, TVC, LCL-filter, and
inner-current-controller dynamics.

\subsubsection{Independent Equilibria and ROA Sections}
\label{subsubsec:gfl_roa_sections}

Although common parameters and external commands are used wherever
possible, the three models do not share an imposed equilibrium. Each
equilibrium is obtained from the corresponding model equations,
\begin{equation}
\boldsymbol{f}_j(\boldsymbol{x}_j^\star)=\boldsymbol{0},
\qquad
j\in\{16,5,3\}.
\label{eq:gfl_independent_equilibria}
\end{equation}
The ODE-based ROA sections are then expressed in model-centered
coordinates,
\begin{equation}
\Delta\delta_j=\delta_j-\delta_j^\star,
\qquad
\Delta\omega_j=\dot\delta_j-\dot\delta_j^\star,
\qquad
j\in\{16,5,3\}.
\label{eq:gfl_centered_coordinates}
\end{equation}
All states not displayed in the two-dimensional section are
initialized at the corresponding model equilibrium, except for the
PLL integral state, which is algebraically adjusted when required to
realize the specified initial value of $\Delta\omega_j$. The models
are not exact model reductions of one another; therefore,
strict nesting of their ROAs is not imposed. Instead, the comparison
evaluates how each modeling layer represents the transient-stability
region and how the model choice affects the conservativeness of the
subsequent SOS-based ROA estimate.

\section{Comparative ROA Assessment}
This section presents a comparative ROA assessment of the GFM and GFL converter models. For both the GFM and GFL model series, the ODE-based ROA estimates obtained from the full- and reduced-order models are evaluated on a
common manifold of admissible initial conditions and compared through
identical two-dimensional state-space slices.

For each reduced-order model, the SOS-certified inner estimate is then compared with ODE-based ROA estimates of both the reduced- and full-order models, allowing its overall discrepancy relative to the full-order reference to be assessed. The GFM and GFL model series are examined separately in the following subsections.

\subsection{GFM Converter Models}

\noindent\textit{1) ODE-Based Assessment of Model-Reduction Effects}

\begin{figure}[H]
\centering
\subfloat[]{\includegraphics[height=4.55cm, trim=45 240 50 255, clip]{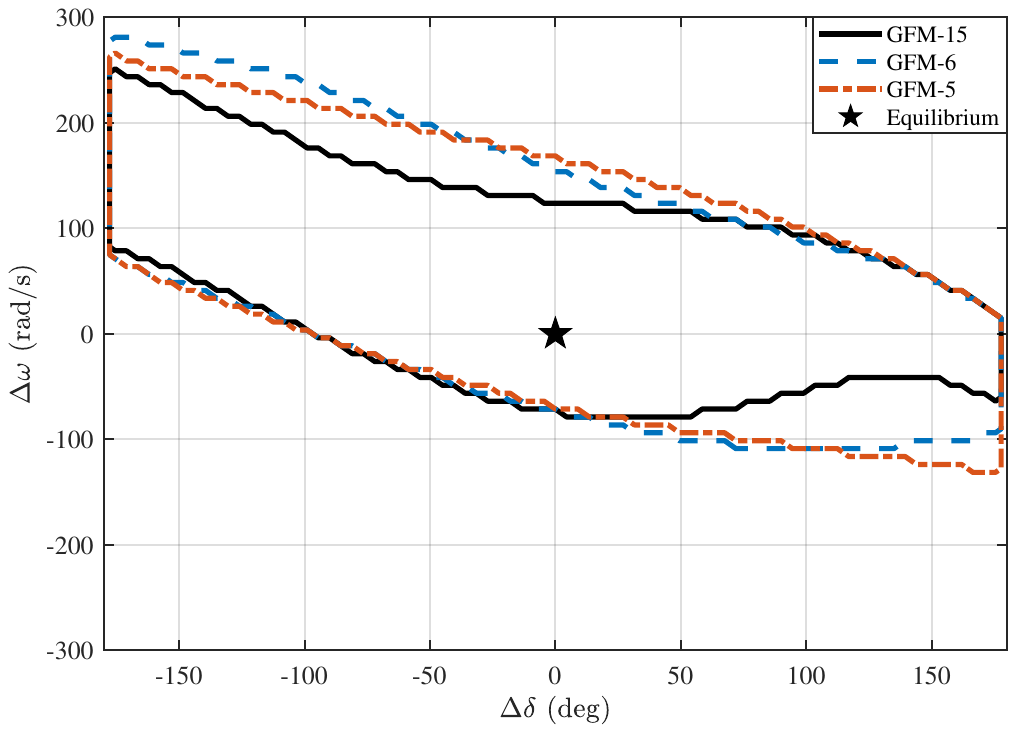}%
\label{fig:gfm_roa_delta_omega}}
\vspace{1mm}
\subfloat[]{\includegraphics[height=4.55cm, trim=45 240 50 255, clip]{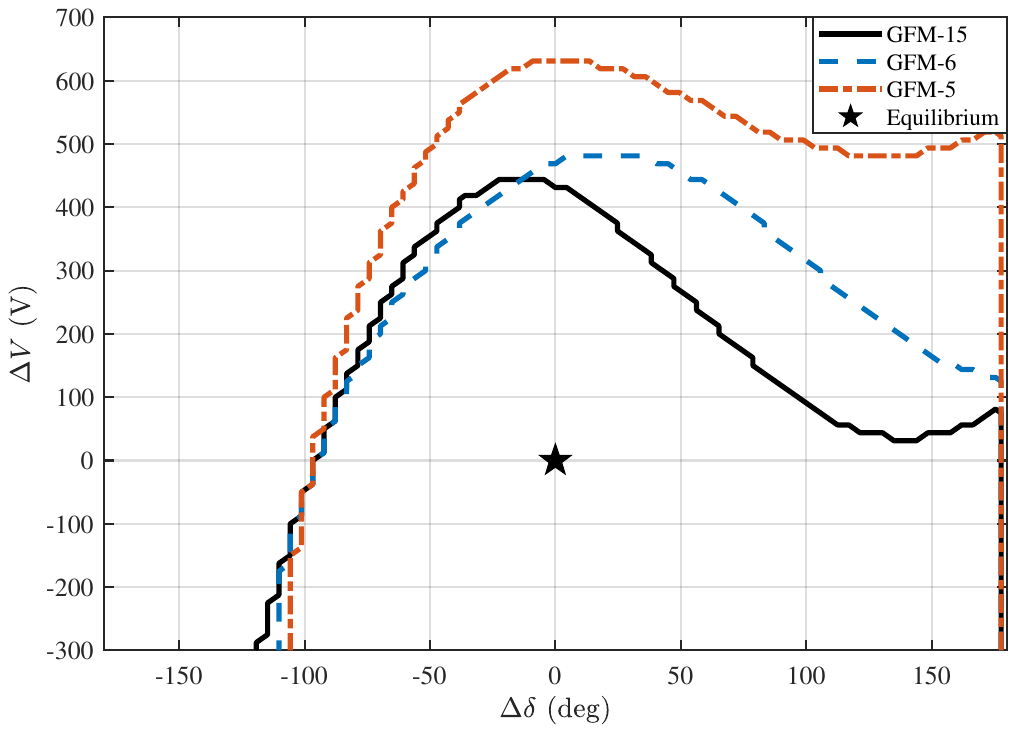}%
\label{fig:gfm_roa_delta_v}}
\caption{Influence of model order on the ODE-based ROA estimates of
the GFM converter models: (a) $\delta$--$\omega$ plane and
(b) $\delta$--$V$ plane.}
\label{fig:gfm_ode_roa}
\end{figure}

To assess the influence of model reduction, Fig.~\ref{fig:gfm_ode_roa} compares the ODE-based ROA cross-sections obtained by time-domain simulation of the nonlinear equations from the three GFM models.

In the $\delta$--$\omega$ plane, all models exhibit similar overall boundary trends, while the reduced-order models generally overestimate the ROA of GFM-15, particularly for positive angle deviations combined with negative frequency deviations. GFM-6 does not provide a clear improvement over GFM-5 in this plane. 

The differences among the models are more substantial in the $\delta$--$V$
plane. GFM-5 substantially overestimates the ROA of GFM-15, whereas
GFM-6 provides a closer approximation by retaining an additional
voltage-related state, although noticeable discrepancies remain at
large positive angle deviations. These results indicate that ROA
accuracy depends not only on model order, but also on whether the
dynamics relevant to the examined state plane are retained.

\noindent\textit{2) Comparison of ODE- and SOS-Based ROA Estimates}

\begin{figure}[H]
\centering
\subfloat[]{%
\includegraphics[
   width=0.7\columnwidth,
   height=4.55cm,
    trim={45 230 40 215},
    clip
]{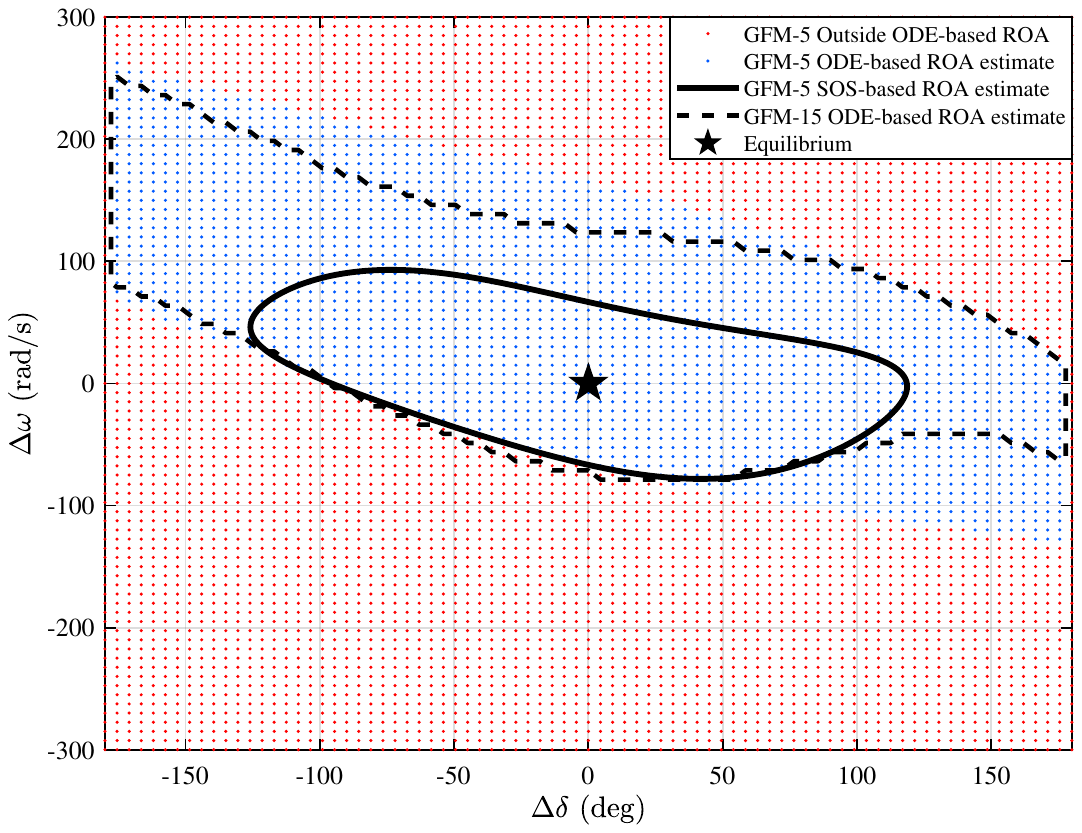}%
\label{fig:gfm5_sos_ode_delta_omega}}
\vspace{1mm}
\subfloat[]{%
\includegraphics[
     width=0.7\columnwidth,
    height=4.55cm,
    trim={45 230 40 215},
    clip
]{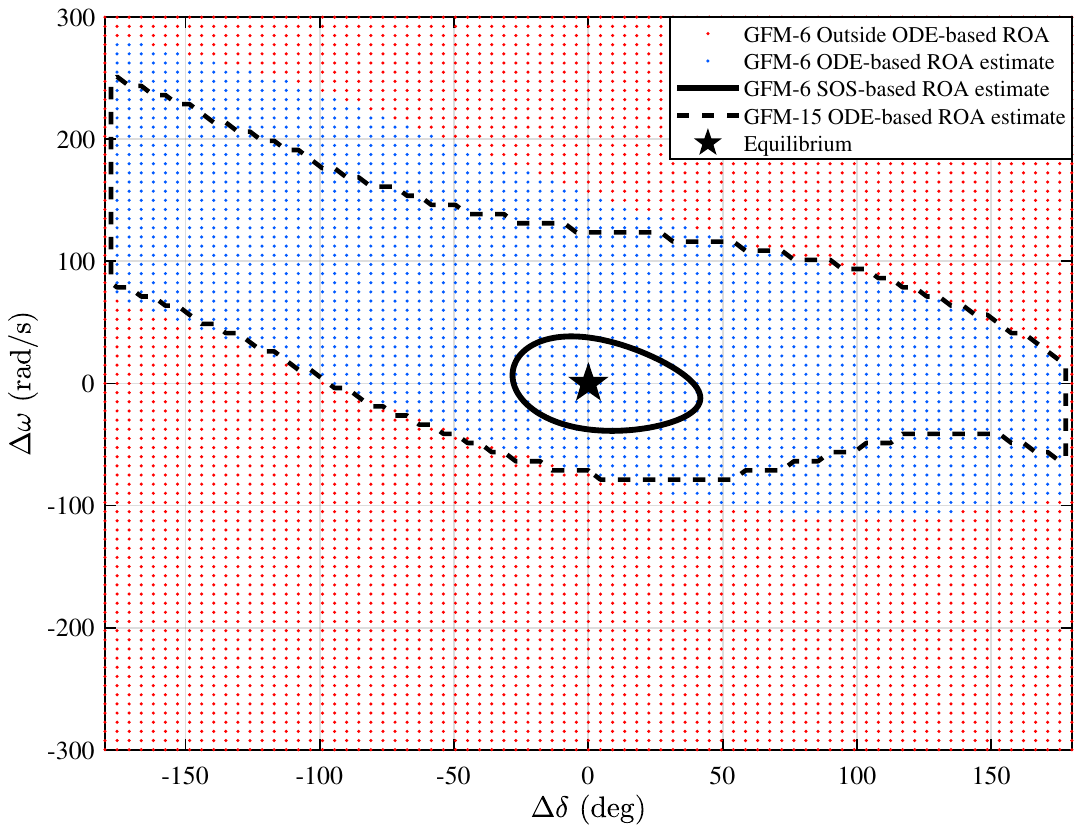}%
\label{fig:gfm6_sos_ode_delta_omega}}
\caption{Comparison of the ODE- and SOS-based ROA estimates in the
$\delta$--$\omega$ plane: (a) GFM-5 and (b) GFM-6.}
\label{fig:gfm_sos_ode_delta_omega}
\end{figure}

\begin{figure}[H]
\centering
\subfloat[]{%
\includegraphics[
     width=0.7\columnwidth,
   height=4.55cm,
    trim={45 230 40 215},
    clip
]{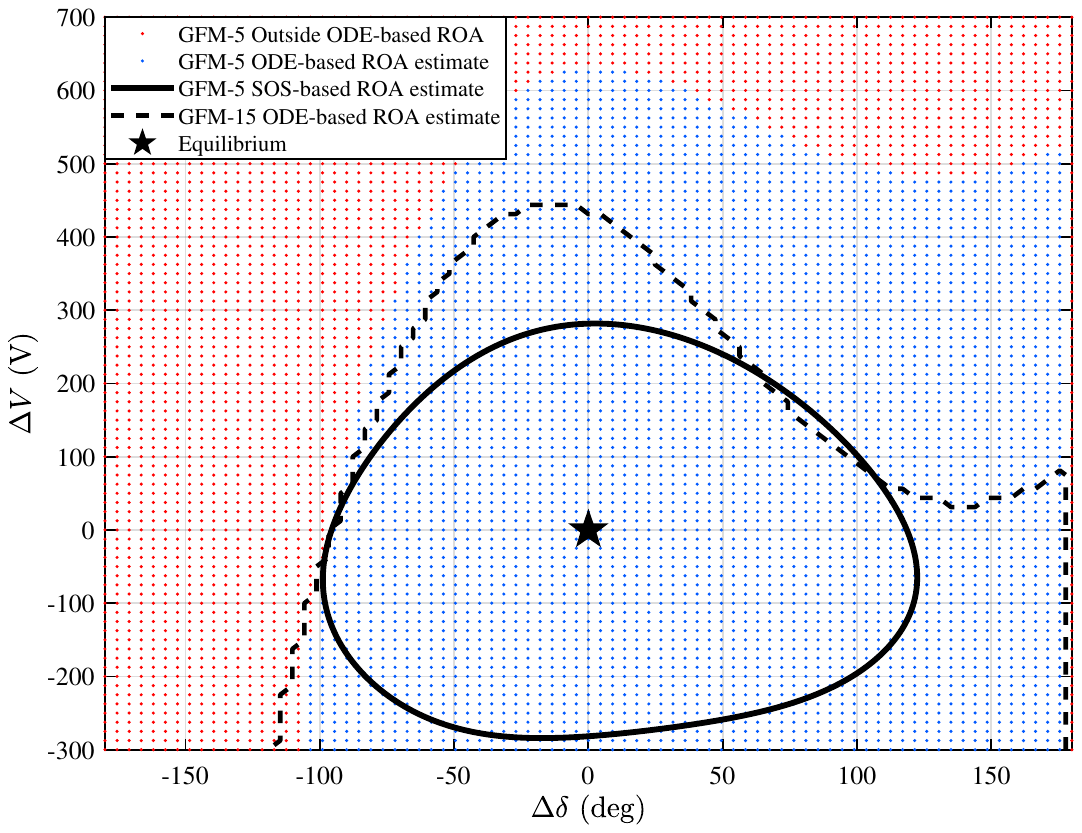}%
\label{fig:gfm5_sos_ode_delta_v}}
\vspace{1mm}
\subfloat[]{%
\includegraphics[
     width=0.7\columnwidth,
   height=4.55cm,
    trim={45 230 40 215},
    clip
]{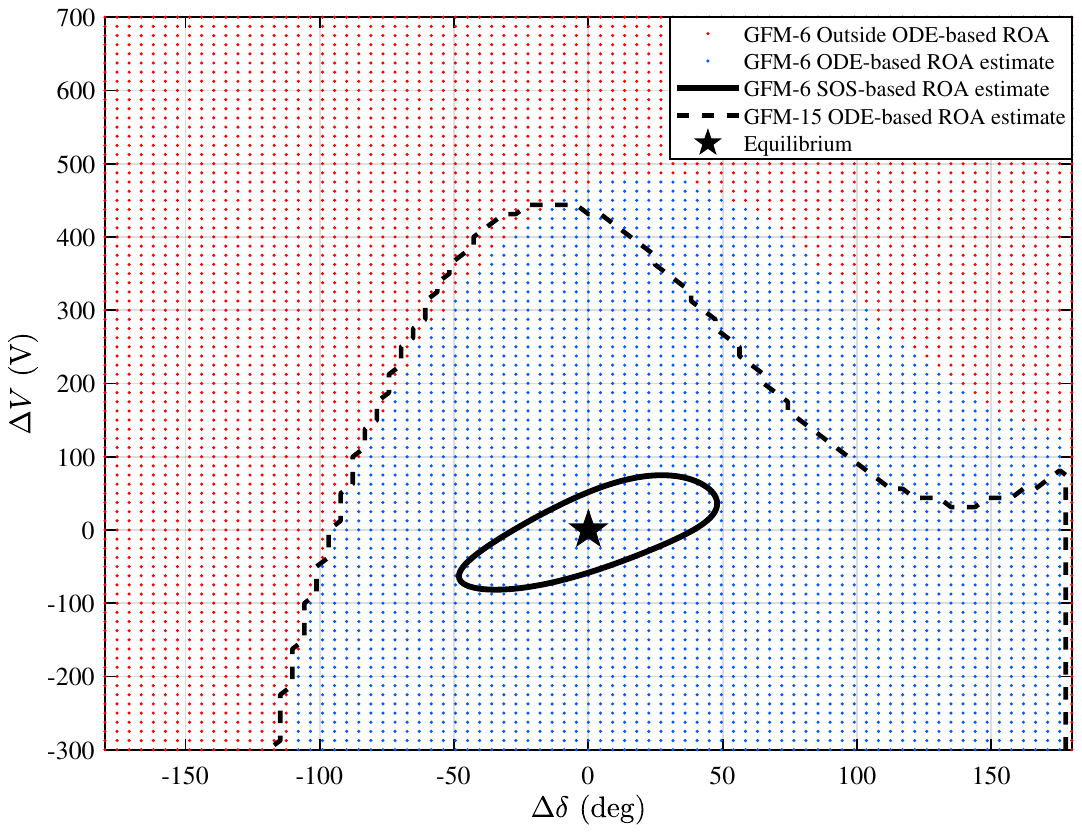}%
\label{fig:gfm6_sos_ode_delta_v}}
\caption{Comparison of the ODE- and SOS-based ROA estimates in the
$\delta$--$V$ plane: (a) GFM-5 and (b) GFM-6.}
\label{fig:gfm_sos_ode_delta_v}
\end{figure}

To assess SOS conservativeness and the overall deviation from the full-order reference, Fig.~\ref{fig:gfm_sos_ode_delta_omega} and Fig.~\ref{fig:gfm_sos_ode_delta_v} compare the SOS estimates of GFM-5 and GFM-6 with the ODE-based ROA estimates of the corresponding reduced-order models and the full-order GFM-15 in the $\delta$--$\omega$ and $\delta$--$V$ planes. Across these slices, the SOS estimates lie largely within the ODE-based ROA estimates of GFM-15.

The comparison in the $\delta$--$\omega$ plane is shown in Fig.~\ref{fig:gfm_sos_ode_delta_omega}. For GFM-5, the SOS estimate captures the negative inclination and principal angle-frequency coupling of the ODE-based ROA, but remains narrower near the upper boundary and at large angle deviations. This gap reflects the difficulty of representing the broad and asymmetric ODE-based region using a quadratic Lyapunov level set. For GFM-6, the SOS estimate contracts substantially while the ODE-based region remains a broad diagonal band. Consequently, increasing model order widens the SOS–ODE boundary gap, particularly along the phase-angle axis and near the frequency limits. Without a corresponding contraction of the ODE-based slice, this indicates greater SOS conservativeness rather than a smaller underlying ROA.

In the $\delta$--$V$ plane (Fig.~\ref{fig:gfm_sos_ode_delta_v}), the GFM-5 ODE-based ROA extends much farther towards positive than negative voltage deviations. The SOS-based estimate closely follows portions of its lower and lateral boundaries near equilibrium but misses the upward extension. For GFM-6, the SOS estimate becomes a smaller, inclined ellipsoidal region, with substantial gaps at large angle deviations and near the upper voltage limit. Increasing model order thus changes the certified region’s geometry without improving agreement, leaving much of the ODE-based high-voltage region uncertified.

The greater conservativeness observed for GFM-6 may be associated with the additional fast voltage state. This state introduces stronger time-scale separation and increases the dimension and complexity of the SOS conditions, making it more difficult for a quadratic Lyapunov function to provide a tight inner approximation of the asymmetric ODE-based ROA.

Overall, increasing the model order does not necessarily improve the agreement between the SOS- and ODE-based ROA estimates. The effect is direction dependent: GFM-6 produces a more restricted SOS certificate in the phase-angle direction, while the voltage-direction extent and orientation of the certified region are modified rather than uniformly reduced.

\subsection{GFL Converter Models}

\noindent\textit{1) ODE-Based Assessment of Model-Reduction Effects}

\begin{figure}[H]
\centering
\includegraphics[width=0.7\columnwidth, height=4.55cm, trim=45 230 70 245, clip]{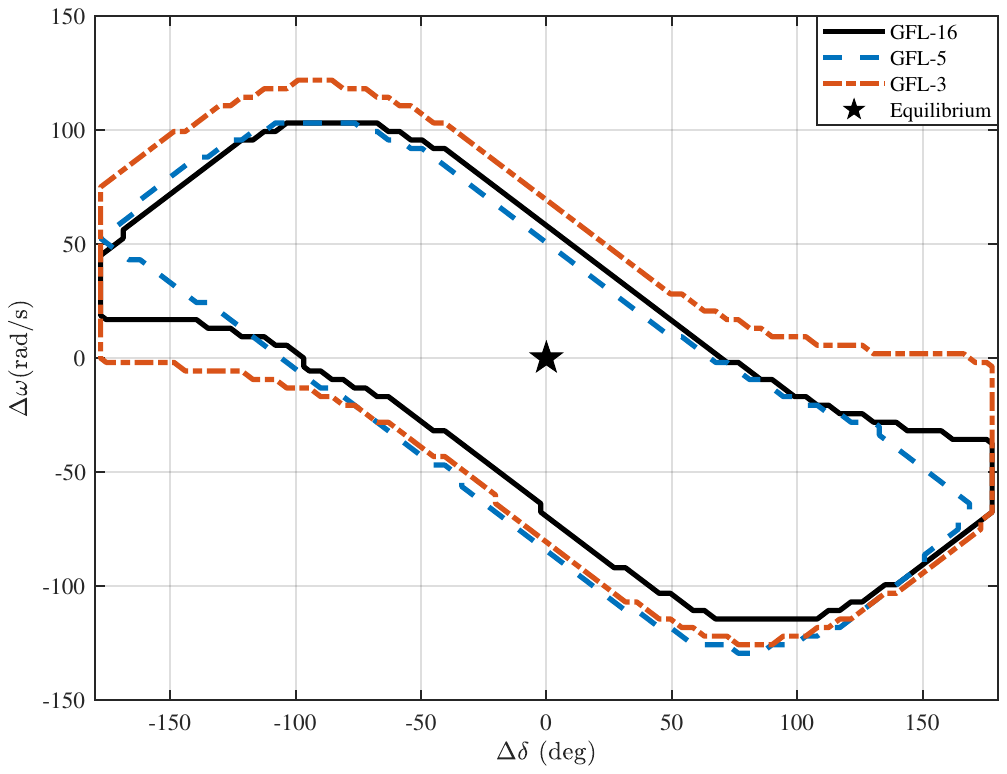}%
\caption{Influence of model order on the ODE-based ROA estimates of
the GFL converter models: $\delta$--$\omega$ plane}
\label{fig:gfl_roa_delta_omega}
\end{figure}

To assess the influence of model reduction, Fig.~\ref{fig:gfl_roa_delta_omega}
compares the ODE-based ROA cross-sections obtained from the three GFL
models.

Along the upper ROA boundary, GFL-5 
and GFL-3 nearly overlap and both lie outside the full-order 
boundary over most of the angle range. Along the lower boundary, GFL-5 
remains close to GFL-16, whereas GFL-3 departs visibly from both, 
particularly for negative frequency deviations, where its ROA encloses 
a substantially larger region. A strict containment of the three ROAs is
not observed: at large positive angle deviations the GFL-16 ROA
locally extends beyond that of GFL-5. Thus, GFL-5 provides a
consistently closer approximation to the full-order ROA than GFL-3,
whose additional simplification of the dc-side dynamics is most
visible in the lower part of this plane.

\noindent\textit{2) Comparison of ODE- and SOS-Based ROA Estimates}

\begin{figure}[H]
\centering
\subfloat[]{\includegraphics[width=0.7\columnwidth, height=4.55cm, trim=45 235 70 245, clip]{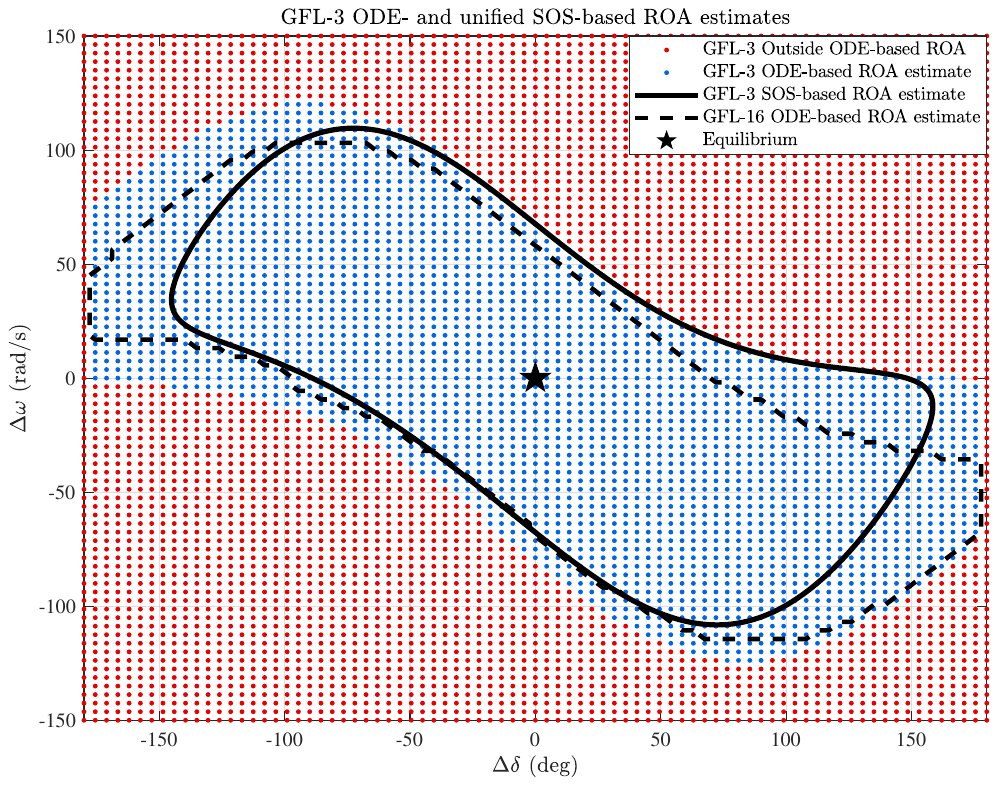}%
\label{fig:gfl3_sos_ode}}
\vspace{1mm}
\subfloat[]{\includegraphics[width=0.7\columnwidth, height=4.55cm, trim=45 235 70 245, clip]{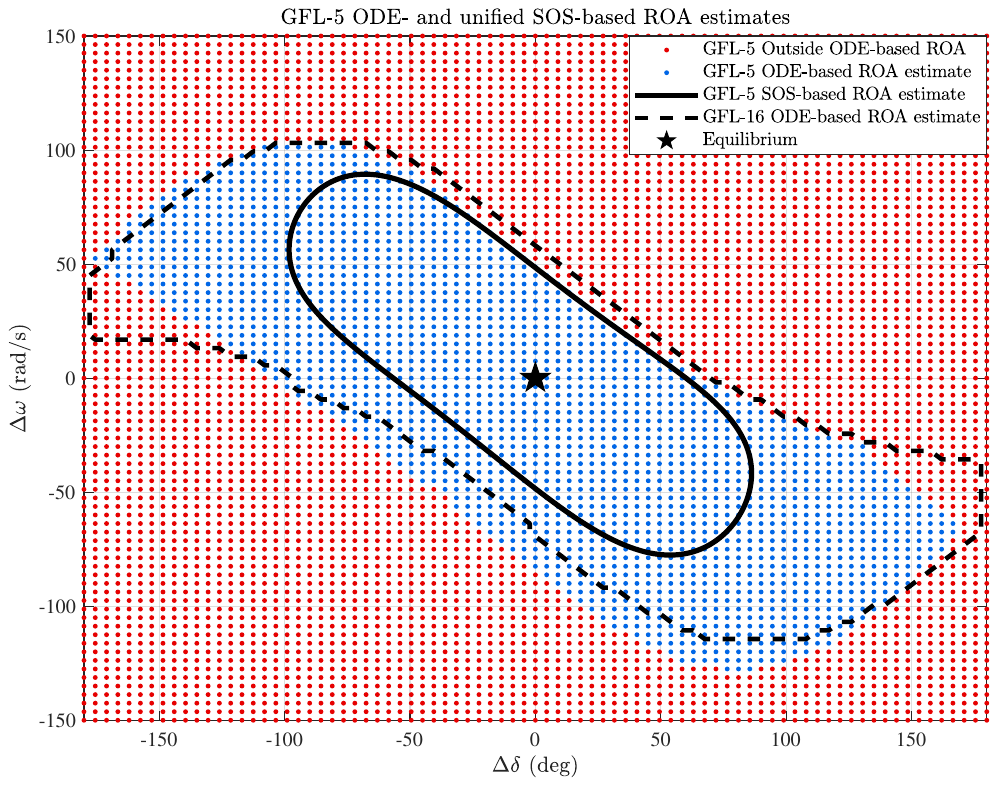}%
\label{fig:gfl5_sos_ode}}
\caption{Comparison of the ODE- and SOS-based ROA estimates for the
reduced-order GFL models: (a) GFL-3 in the $\delta$--$\omega$ plane;
(b) GFL-5 in the $\delta$--$\omega$ plane.}
\label{fig:gfl_sos_ode_comparison}
\end{figure}

The conservativeness of the SOS-certified ROAs for the two reduced-order GFL models, GFL-3 and GFL-5, is evaluated by comparison with their corresponding ODE-based reference ROAs, with the ODE-based ROA of the full-order GFL-16 model serving as an additional reference. Fig.~\ref{fig:gfl_sos_ode_comparison} presents the comparisons in the $\delta$--$\omega$ plane. The GFL-3 SOS estimate extends beyond portions of the GFL-16 ODE-based boundary, whereas the GFL-5 estimate lies largely within it.

Fig.~\ref{fig:gfl_sos_ode_comparison} compares the SOS-certified ROA with the corresponding ODE-based ROA estimates of the GFL
models. For GFL-3, the SOS-certified boundary captures the orientation
and closely follows a large portion of the ODE-based boundary. For
GFL-5, the main directional trend is retained, but the certified region
has a noticeably smaller phase-angle span and remains farther from the
ODE-based boundary.

A key observation is that retaining additional converter dynamics does
not necessarily improve the SOS-based ROA estimate and may instead
increase its conservativeness. This may be because the additional states
increase the dimension of the SOS problem and require the Lyapunov and
SOS conditions to hold over a higher-dimensional state space, making it
more difficult to certify a large region.

\begin{table*}[t]
\centering
\caption{ROA coverage, classification discrepancies, and computational cost}
\label{tab:roa_accuracy_cost}
\small
\setlength{\tabcolsep}{5pt}
\renewcommand{\arraystretch}{1.15}

\begin{tabular}{llccccccc}
\toprule
& & \multicolumn{5}{c}{ROA metrics (\%)}
  & \multicolumn{2}{c}{Computation time (s)} \\
\cmidrule(lr){3-7}
\cmidrule(lr){8-9}
Model
& Slice
& $C_{\mathrm{ROM}}$
& $\mathrm{FP}_{\mathrm{ROM}}$
& $C_{\mathrm{SOS},F}$
& $C_{\mathrm{SOS},R}$
& $\mathrm{FP}_{\mathrm{SOS}}$
& $t_{\mathrm{SOS}}$
& $t_{\mathrm{ODE}}$ \\
\midrule

GFM-5
& $\delta$--$\omega$
& \textit{98.968} & \textit{22.66} & \textit{44.069}
& \textit{34.617} & \textit{0.51746} & \textit{57.025} & \textit{42.510} \\

GFM-5
& $\delta$--$V$
& \textit{99.092} & \textit{34.418} & \textit{63.729}
& \textit{42.478} & \textit{0.70713} & \textit{shared} & \textit{121.215} \\

GFM-6
& $\delta$--$\omega$
& \textit{99.083} & \textit{22.501} & \textit{7.0487}
& \textit{5.5132} & \textit{0} & \textit{96.544} & \textit{42.660} \\

GFM-6
& $\delta$--$V$
& \textit{98.147} & \textit{16.398} & \textit{5.4841}
& \textit{4.6714} & \textit{0} & \textit{shared} & \textit{62.841} \\

\midrule

GFL-3
& $\delta$--$\omega$
& \textit{100} & \textit{24.459} & \textit{82.996}
& \textit{71.21} & \textit{11.956} & \textit{24.599} & \textit{8.275} \\

GFL-5
& $\delta$--$\omega$
& \textit{91.201} & \textit{10} & \textit{43.5}
& \textit{42.927} & \textit{0} & \textit{80.199} & \textit{9.255} \\

\midrule
\multicolumn{9}{l}{\textit{Full-order reference models}} \\

GFM-15
& $\delta$--$\omega$
& --- & --- & --- & --- & --- & --- & \textit{83.242} \\

GFM-15
& $\delta$--$V$
& --- & --- & --- & --- & --- & --- & \textit{130.011} \\

GFL-16
& $\delta$--$\omega$
& --- & --- & --- & --- & --- & --- & \textit{530.664} \\

\bottomrule
\end{tabular}

\vspace{2mm}
\begin{minipage}{0.98\textwidth}
\footnotesize
\textit{Metric definitions:}
Let $F$ and $R$ denote the full-order and reduced-order
ODE-based reference regions, respectively, and let $S$ denote
the reduced-order SOS estimate on the corresponding slice.
The reported percentages are
\[
\begin{aligned}
C_{\mathrm{ROM}}
  &=100\frac{|R\cap F|}{|F|},
&
\mathrm{FP}_{\mathrm{ROM}}
  &=100\frac{|R\setminus F|}{|R|},
&
C_{\mathrm{SOS},F}
  &=100\frac{|S\cap F|}{|F|},
&
C_{\mathrm{SOS},R}
  &=100\frac{|S\cap R|}{|R|},
&
\mathrm{FP}_{\mathrm{SOS}}
  &=100\frac{|S\setminus F|}{|S|}.
\end{aligned}
\]

\smallskip
\textit{Computing environment:}
CPU: \textit{[i9-13980HX]};
RAM: \textit{[64G ddr5]};
MATLAB: \textit{[24.2.0.3212159 (R2024b) Update 9]};
YALMIP: \textit{[20250626]};
MOSEK: \textit{[11.0.18]}.
Note that \(t_{\mathrm{ODE}}\) is only the time required to calculated a 2D slide of the ROA, where \(t_{\mathrm{SOS}}\) allows the calculation of the full-order (for that model) ROA. Hence, they should not be directly compared.
\end{minipage}
\end{table*}

\subsection{Comparative Discussion}

The ODE comparisons show that GFL-5 better matches the full-order boundary than GFL-3 over much of the \(\delta\)–\(\omega\) plane, while GFM-6 better captures the full-order voltage boundary than GFM-5 in the \(\delta\)–\(V\) plane. However, both model families exhibit reduced-order ROA portions outside the full-order reference, indicating that model reduction does not consistently produce conservative estimates.

Closer ODE agreement does not ensure a larger SOS-certified ROA: GFL-3 yields a larger certificate than GFL-5, and GFM-5 exceeds GFM-6 in the \(\delta\)–\(\omega\) plane. In the \(\delta\)–\(V\) plane, the GFM-6 SOS estimate spans narrower voltage and angle ranges than GFM-5. Table~\ref{tab:roa_accuracy_cost} quantifies these trade-offs in terms of ROA coverage, false-positive proportions, and computational cost, showing that the higher-order reduced models require longer SOS computation times without achieving greater coverage of the full-order reference ROA. Model selection should therefore account for both full-order ROA fidelity and SOS coverage of the required operating range, including its directional dependence.

\section{Practical Guidelines for Model Selection}
\subsection{Model Selection for SOS-Based ROA Certification}

Model selection for SOS-based ROA estimation should consider both the size of the certified region and its agreement with the full-order ODE reference. A larger SOS-certified ROA is desirable when it captures more of the stable region identified by the full-order model. Increasing model order is justified when it improves this agreement or reduces the gap between the SOS-certified and ODE-based ROAs of the reduced-order model.

\subsection{Recommended Model Choices}

For GFL, the third-order model provides broader SOS-certified coverage, while the fifth-order model better matches much of the full-order ODE boundary. Model selection should therefore consider how well each certificate covers the full-order reference ROA, rather than its size alone.

For GFM, the fifth-order model is preferred for broader angle–frequency certification. The sixth-order model is an improvement for frequency response matching but is not useful for ROA certification, whereas the fifth-order model provides wider certified angle coverage.

Overall, lower-order models are effective candidates for SOS-based ROA estimation. Additional states are justified when they help the SOS-certified region cover more of the full-order reference ROA or reduce overestimation relative to that reference. Closer agreement between ODE-based boundaries alone does not establish a better model for SOS certification.

\section{Conclusion}

This study shows that model accuracy and SOS certification performance are distinct criteria for reduced-order converter models. Increasing model order improves neither automatically. The value of a reduced-order model for SOS analysis lies in how effectively its certificate represents the full-order stability region, rather than in the number of retained states. 

In practical engineering applications, selecting reduced-order models only from waveform or frequency-response fidelity can lead to overly  conservative stability certificates or an inaccurate representation of the full-order stability boundary. In particular, a highly conservative SOS-based ROA estimate may lead to unnecessarily restrictive operating constraints, limiting the effective use of the system’s available recovery capability. These findings provide a basis for balancing model complexity, stability-boundary fidelity, and the conservatism of SOS-based ROA estimates for both GFM and GFL converters.                                                       
\bibliographystyle{IEEEtran}
\bibliography{references}

\clearpage

\appendices

\section{Implementation of the SOS Algorithm}
\label{SOS_Algorithm}

This appendix presents the implementation details of the alternating SOS algorithm, including initialization, Lyapunov and multiplier updates, and termination criteria.
\begin{enumerate}

\item \textbf{Polynomial reformulation.}
Shift the equilibrium to the origin and normalize the state variables. Introduce polynomial auxiliary variables for the trigonometric terms and retain the associated equality $g(x)=0$. For rational models, clear the denominator within the domain defined by $h(x)\geq0$.

\item \textbf{Initialization.}
Let $A$ be the Jacobian of the transformed polynomial vector field at the origin. Let $N$ contain an orthonormal basis for the tangent space $\ker Dg(0)$, and let $U$ contain an orthonormal basis for its orthogonal complement. The tangent linearization is $A_{\mathrm{tan}}=N^{\top}AN$. After verifying that $A_{\mathrm{tan}}$ is Hurwitz, construct the stable extension

\begin{equation}
\begin{aligned}
A_{\mathrm{init}}
&= NA_{\mathrm{tan}}N^{\top}-\alpha UU^{\top},\\
\alpha
&= \max\{1,\|A_{\mathrm{tan}}\|_2\}.
\end{aligned}
\label{eq:sos_init_matrix}
\end{equation}

Solve the Lyapunov equation
\begin{equation}
A_{\mathrm{init}}^{\top}P
+PA_{\mathrm{init}}=-I
\label{eq:sos_init_lyapunov}
\end{equation}
and set $V_{\mathrm{init}}(x)=x^{\top}Px$. With $V_{\mathrm{init}}$ fixed, use a bracket search and bisection to find a feasible level $\lambda_0$ satisfying the Lyapunov-decrease and domain-containment conditions. Normalize the candidate as $V_0=V_{\mathrm{init}}/\lambda_0$ and fix $\lambda=1$ for the remaining iterations. A fixed-$V_0$ feasibility search then provides the initial $\beta$ and multipliers for the normalized constraint set.

\item \textbf{Lyapunov step.}
Fix the principal multipliers $s_1$, $s_2$, and $s_3$
from the preceding feasible solution.
Optimize the coefficients of $V$ and maximize $\beta$
subject to the positivity condition,
\eqref{eq:lifted_sos_constraints}, and
\eqref{eq:sos_domain_containment}, with $\lambda=1$.
The equality multipliers $\ell_1,\ell_2,\ell_3$ and
the domain multipliers $\sigma_1,\sigma_2$ are updated
in this step.
The implemented search uses a homogeneous quadratic
polynomial $V$ in the transformed coordinates.

\item \textbf{Multiplier step.}
Fix the updated $V$ and search for a larger feasible
$\beta$ by bracket expansion and bisection.
At each trial value, solve an SDP feasibility problem
for $s_1,s_2,s_3$, $\sigma_1,\sigma_2$, and
$\ell_1,\ell_2,\ell_3$.
This step verifies the updated Lyapunov function and
provides the multipliers for the next iteration.

\item \textbf{Convergence check.}
Accept the updated solution if feasibility is verified
and $\beta$ does not decrease within the numerical
acceptance tolerance.
Repeat steps 3--4 until the relative improvement in
$\beta$ falls below the prescribed tolerance or the
maximum iteration count is reached.
If an update fails, retain the last accepted feasible
solution.
Return
$\widehat{\mathcal{R}}
=\{x\in\mathcal{M}:V^{\star}(x)<1\}$.

\end{enumerate}

In the implementation, $\rho(x)=x^{\top}x$ is defined
in the normalized transformed coordinates, and
$\varepsilon_1=\varepsilon_2=10^{-6}$.
The polynomial multipliers have degree at most two.
The relative search and alternating-iteration
tolerances are both $5\times10^{-3}$, and the maximum
number of alternating iterations is 40.
State normalization improves numerical scaling,
while tangent-space initialization accounts for the
algebraic structure of the transformed models.

\section{Complete Full-Order GFM-15 Model}
\label{app:gfm_full_model}

Converter-side quantities, denoted by the subscript $dq$, are expressed
in the local frame rotating at the converter frequency $\omega$.
Network quantities, denoted by the subscript $DQ$, are expressed in the
synchronous frame rotating at the nominal frequency $\omega_0$, whose
$D$-axis is aligned with the stiff-grid voltage. The angle between the
two frames is $\delta$. Define
\begin{equation}
\mathcal{T}(\delta)=
\begin{bmatrix}
\cos\delta&-\sin\delta\\
\sin\delta&\phantom{-}\cos\delta
\end{bmatrix},
\quad
J=\begin{bmatrix}0&1\\-1&0\end{bmatrix},
\quad
e=\begin{bmatrix}1\\0\end{bmatrix},
\label{eq:gfm15_notation}
\end{equation}
where $I_2$ denotes the $2\times2$ identity matrix.

\subsection{Droop and Power Dynamics}

The angle, frequency-droop, and voltage-droop dynamics are
\begin{subequations}
\label{eq:gfm15_droop}
\begin{align}
\dot{\delta}
&=\omega-\omega_0,
\label{eq:gfm15_delta}\\
\tau\dot{\omega}
&=-\omega+\omega_n-k_pP,
\label{eq:gfm15_omega}\\
\tau\dot{V}
&=-V+V_n-k_qQ.
\label{eq:gfm15_V}
\end{align}
\end{subequations}
Here, $V$ is the voltage command generated by the droop controller,
$(\omega_n,V_n)$ are the no-load frequency and voltage set points,
and $(k_p,k_q)$ are the active- and reactive-power droop gains. The
injected powers are
\begin{subequations}
\label{eq:gfm15_PQ}
\begin{align}
P
&=\boldsymbol{i}_{odq}^{\mathsf T}\boldsymbol{v}_{odq},
\label{eq:gfm15_P}\\
Q
&=\boldsymbol{i}_{odq}^{\mathsf T}J\boldsymbol{v}_{odq}.
\label{eq:gfm15_Q}
\end{align}
\end{subequations}

\subsection{Voltage and Current Controllers}

The outer voltage controller regulates the filter-capacitor voltage
$\boldsymbol{v}_{odq}$ to the command $eV$ and generates the current
reference. The inner current controller generates the converter
terminal-voltage command:
\begin{subequations}
\label{eq:gfm15_control}
\begin{align}
\dot{\boldsymbol{\phi}}_{dq}
&=eV-\boldsymbol{v}_{odq},
\label{eq:gfm15_voltage_integrator}\\
\boldsymbol{i}_{dq}^{\rm ref}
&=K_{PV}\left(eV-\boldsymbol{v}_{odq}\right)
+K_{IV}\boldsymbol{\phi}_{dq},
\label{eq:gfm15_vloop}\\
\dot{\boldsymbol{\gamma}}_{dq}
&=\boldsymbol{i}_{dq}^{\rm ref}-\boldsymbol{i}_{dq},
\label{eq:gfm15_current_integrator}\\
\boldsymbol{v}_{dq}
&=\boldsymbol{v}_{odq}
+K_{PC}\left(\boldsymbol{i}_{dq}^{\rm ref}
-\boldsymbol{i}_{dq}\right)
+K_{IC}\boldsymbol{\gamma}_{dq}
-\omega L_fJ\boldsymbol{i}_{dq}.
\label{eq:gfm15_cloop}
\end{align}
\end{subequations}
The controller states are
$\boldsymbol{\phi}_{dq},\boldsymbol{\gamma}_{dq}\in\mathbb{R}^2$.
The output-voltage feedforward and the decoupling term
$-\omega L_fJ\boldsymbol{i}_{dq}$ follow the standard
synchronous-frame current-controller structure \cite{6200347}.

\subsection{LC Filter and Network Dynamics}

The local-frame filter-inductor and filter-capacitor dynamics are
\begin{subequations}
\label{eq:gfm15_filter}
\begin{align}
L_f\dot{\boldsymbol{i}}_{dq}
&=\omega L_fJ\boldsymbol{i}_{dq}
-R_f\boldsymbol{i}_{dq}
+\boldsymbol{v}_{dq}-\boldsymbol{v}_{odq},
\label{eq:gfm15_Lf}\\
C_f\dot{\boldsymbol{v}}_{odq}
&=\omega C_fJ\boldsymbol{v}_{odq}
+\boldsymbol{i}_{dq}-\boldsymbol{i}_{odq}.
\label{eq:gfm15_Cf}
\end{align}
\end{subequations}
In the synchronous frame, the $RL$ line connecting the converter to
the stiff source and the local $RL$ load satisfy
\begin{subequations}
\label{eq:gfm15_network}
\begin{align}
L_{\ell}\dot{\boldsymbol{I}}_{DQ}^{\ell}
&=\left(-R_{\ell}I_2+\omega_0L_{\ell}J\right)
\boldsymbol{I}_{DQ}^{\ell}
+\mathcal{T}(\delta)\boldsymbol{v}_{odq}-eV_0,
\label{eq:gfm15_line}\\
L_L\dot{\boldsymbol{I}}_{DQ}^{L}
&=\left(-R_LI_2+\omega_0L_LJ\right)
\boldsymbol{I}_{DQ}^{L}
+\mathcal{T}(\delta)\boldsymbol{v}_{odq}.
\label{eq:gfm15_load}
\end{align}
\end{subequations}
The total output current expressed in the converter frame is
\begin{equation}
\boldsymbol{i}_{odq}
=\mathcal{T}^{-1}(\delta)
\left(\boldsymbol{I}_{DQ}^{\ell}
+\boldsymbol{I}_{DQ}^{L}\right).
\label{eq:gfm15_io}
\end{equation}
Equations~\eqref{eq:gfm15_droop}--\eqref{eq:gfm15_io} constitute
the complete GFM-15 benchmark.

\section{Frequency-Response Comparison of the GFM Models}
\label{app:gfm6_frequency}

\begin{figure}[H]
    \centering
    \includegraphics[width=\columnwidth, trim=40 220 40 240, clip]{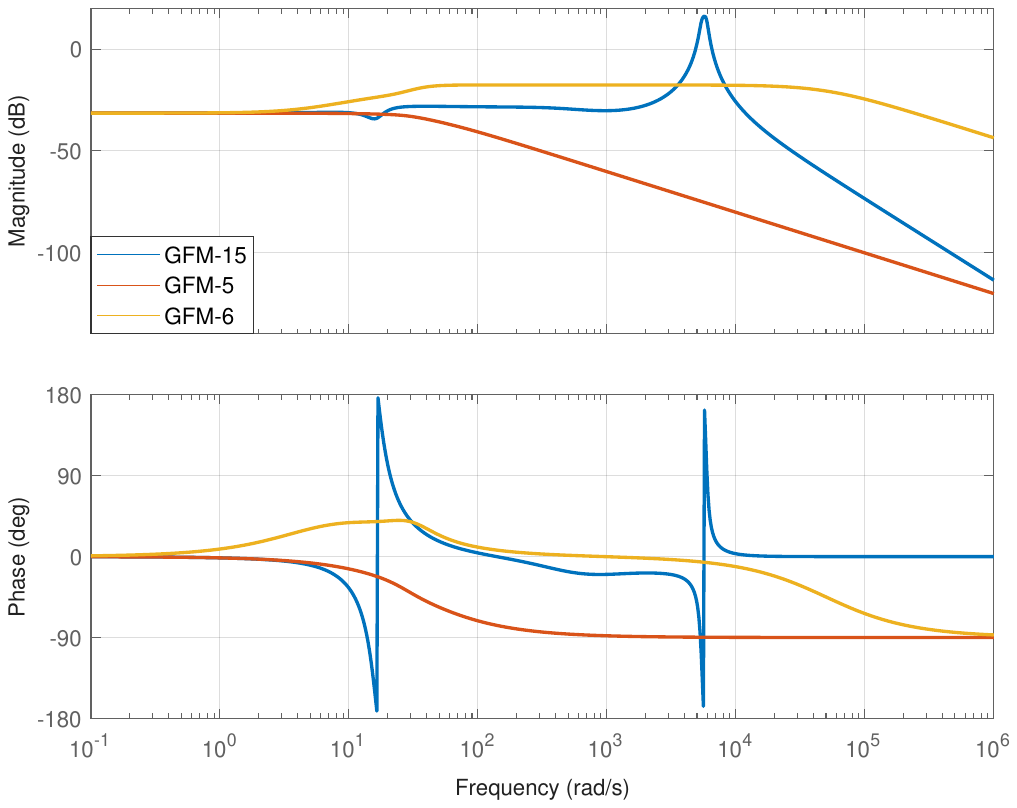}
    \caption{Frequency-response comparison of GFM-15, GFM-5,
    and GFM-6 at the same operating point.}
    \label{fig:gfm_frequency}
\end{figure}

Fig.~\ref{fig:gfm_frequency} compares the frequency responses of
GFM-15, GFM-5, and GFM-6 linearized at the same operating point.
Compared with GFM-5, GFM-6 better captures the gain increase and
phase-lead characteristic of GFM-15 in the intermediate-frequency
range, supporting the inclusion of the additional voltage-related
state. However, the high-frequency resonance is not reproduced,
and the improved agreement is limited to part of the frequency range.

\section{Complete Full-Order GFL-16 Model}
\label{app:gfl_full_model}

The $DQ$ frame is synchronized with the grid and its $D$-axis is
aligned with the grid voltage, whereas the $dq$ frame is synchronized
with the PLL. With the relative-angle convention in
\eqref{eq:gfl16_angle_definition}, define
\begin{equation}
\boldsymbol{T}_{\rm pll}(\vartheta)=
\begin{bmatrix}
\cos\vartheta&\sin\vartheta\\
-\sin\vartheta&\cos\vartheta
\end{bmatrix}.
\label{eq:gfl16_transform}
\end{equation}
Because the input variables are already expressed in the
grid-synchronous $DQ$ frame, the transformation to the PLL frame
depends only on the relative angle:
\begin{subequations}
\label{eq:gfl16_frame_variables}
\begin{align}
\begin{bmatrix}
v_{cd}^{\rm pll}\\v_{cq}^{\rm pll}
\end{bmatrix}
&=\boldsymbol{T}_{\rm pll}(-\delta_{16})
\begin{bmatrix}
v_{cD}\\v_{cQ}
\end{bmatrix},
\label{eq:gfl16_frame_voltage}\\
\begin{bmatrix}
i_{1d}^{\rm pll}\\i_{1q}^{\rm pll}
\end{bmatrix}
&=\boldsymbol{T}_{\rm pll}(-\delta_{16})
\begin{bmatrix}
i_{1D}\\i_{1Q}
\end{bmatrix}.
\label{eq:gfl16_frame_current}
\end{align}
\end{subequations}

\subsection{Reactive-Power and Outer-Control Dynamics}

The reactive power and its measured value satisfy
\begin{subequations}
\label{eq:gfl16_reactive_power}
\begin{align}
Q
&=\frac{3}{2}\left(
v_{cd}^{\rm pll}i_{1q}^{\rm pll}
-v_{cq}^{\rm pll}i_{1d}^{\rm pll}\right),
\label{eq:gfl16_Q}\\
T_m\dot Q_f
&=Q-Q_f.
\label{eq:gfl16_Qf}
\end{align}
\end{subequations}
The dc-voltage and reactive-power controllers generate the current
references
\begin{subequations}
\label{eq:gfl16_current_references}
\begin{align}
i_d^{\rm ref}
&=K_{pv}\left(V_{\rm dc}-V_{\rm dc}^{\rm ref}\right)
+\frac{\xi_v}{T_{iv}},
\label{eq:gfl16_id_ref}\\
i_q^{\rm ref}
&=K_{pq}\left(Q^{\rm ref}-Q_f\right)
+\frac{\xi_q}{T_{iq}},
\label{eq:gfl16_iq_ref}
\end{align}
\end{subequations}
with the integrator dynamics
\begin{subequations}
\label{eq:gfl16_controller_integrators}
\begin{align}
\dot{\xi}_{id}
&=i_d^{\rm ref}-i_{1d}^{\rm pll},
\label{eq:gfl16_xid}\\
\dot{\xi}_{iq}
&=i_q^{\rm ref}-i_{1q}^{\rm pll},
\label{eq:gfl16_xiq}\\
\dot{\xi}_v
&=V_{\rm dc}-V_{\rm dc}^{\rm ref},
\label{eq:gfl16_xv}\\
\dot{\xi}_q
&=Q^{\rm ref}-Q_f.
\label{eq:gfl16_xq}
\end{align}
\end{subequations}

\subsection{Inner Current Controller and PWM Delay}

The inner current-controller outputs in the PLL frame are
\begin{subequations}
\label{eq:gfl16_current_controller}
\begin{align}
E_{1d}
&=K_{pi}\left(i_d^{\rm ref}-i_{1d}^{\rm pll}\right)
+\frac{\xi_{id}}{T_{ii}}
+\omega_0L_1i_{1q}^{\rm pll}
+v_{cd}^{\rm pll},
\label{eq:gfl16_E1d}\\
E_{1q}
&=K_{pi}\left(i_q^{\rm ref}-i_{1q}^{\rm pll}\right)
+\frac{\xi_{iq}}{T_{ii}}
-\omega_0L_1i_{1d}^{\rm pll}
+v_{cq}^{\rm pll}.
\label{eq:gfl16_E1q}
\end{align}
\end{subequations}
Their grid-frame references and the PWM-delay dynamics are
\begin{subequations}
\label{eq:gfl16_pwm}
\begin{align}
\begin{bmatrix}
E_D^{\rm ref}\\E_Q^{\rm ref}
\end{bmatrix}
&=\boldsymbol{T}_{\rm pll}^{-1}(-\delta_{16})
\begin{bmatrix}
E_{1d}\\E_{1q}
\end{bmatrix},
\label{eq:gfl16_voltage_transform}\\
T_d\dot E_D
&=E_D^{\rm ref}-E_D,
\label{eq:gfl16_Ed}\\
T_d\dot E_Q
&=E_Q^{\rm ref}-E_Q.
\label{eq:gfl16_Eq}
\end{align}
\end{subequations}

\subsection{LCL-Filter and DC-Link Dynamics}

The LCL-filter dynamics in the grid-synchronous frame are
\begin{subequations}
\label{eq:gfl16_lcl}
\begin{align}
L_1\dot i_{1D}
&=E_D-v_{cD}-\omega_0L_1i_{1Q},
\label{eq:gfl16_i1D}\\
L_1\dot i_{1Q}
&=E_Q-v_{cQ}+\omega_0L_1i_{1D},
\label{eq:gfl16_i1Q}\\
C_f\dot v_{cD}
&=i_{1D}-i_{2D}-\omega_0C_fv_{cQ},
\label{eq:gfl16_vcD}\\
C_f\dot v_{cQ}
&=i_{1Q}-i_{2Q}+\omega_0C_fv_{cD},
\label{eq:gfl16_vcQ}\\
L_2\dot i_{2D}
&=v_{cD}-u_{gD}-\omega_0L_2i_{2Q},
\label{eq:gfl16_i2D}\\
L_2\dot i_{2Q}
&=v_{cQ}-u_{gQ}+\omega_0L_2i_{2D}.
\label{eq:gfl16_i2Q}
\end{align}
\end{subequations}
The converter power and dc-link dynamics are
\begin{subequations}
\label{eq:gfl16_dc_subsystem}
\begin{align}
P_c
&=\frac{3}{2}\left(E_Di_{1D}+E_Qi_{1Q}\right),
\label{eq:gfl16_power}\\
C_{\rm dc}\dot V_{\rm dc}
&=I_{\rm src}-\frac{P_c}{V_{\rm dc}}.
\label{eq:gfl16_vdc}
\end{align}
\end{subequations}

\subsection{PLL Dynamics}

The normalized PLL equations associated with the positive-angle
coordinate in \eqref{eq:gfl16_angle_definition} are
\begin{subequations}
\label{eq:gfl16_pll}
\begin{align}
\Delta\omega_{16}=\dot{\delta}_{16}
&=\frac{-K_{p,{\rm pll}}v_{cq}^{\rm pll}
+K_{i,{\rm pll}}\xi_{\rm pll}}{V_g},
\label{eq:gfl16_delta}\\
\dot{\xi}_{\rm pll}
&=-v_{cq}^{\rm pll}.
\label{eq:gfl16_pll_integrator}
\end{align}
\end{subequations}
Equations~\eqref{eq:gfl16_frame_variables}--\eqref{eq:gfl16_pll},
constitute the complete GFL-16 benchmark.

\newpage

\end{document}